\documentclass[10pt,a4paper]{article}

\usepackage{jcappub}
\usepackage{amsmath}
\usepackage{amssymb}

\usepackage{graphicx}
\usepackage{amssymb}
\usepackage{amsmath}

\usepackage{color}
\newcommand{\cred}{\color{black}}
\newcommand{\be}{\begin{equation}}
\newcommand{\ee}{\end{equation}}
\newcommand{\bea}{\begin{eqnarray}}
\newcommand{\eea}{\end{eqnarray}}

\title{Bulk Viscous Matter-dominated Universes: Asymptotic Properties}

\author[a]{Arturo Avelino,}
\author[b]{Ricardo Garc\'{\i}a-Salcedo,}
\author[c]{Tame Gonzalez,}
\author[d]{Ulises Nucamendi,}
\author[e]{and Israel Quiros}

\affiliation[a]{Departamento de F\'isica, Campus Le\'on, Universidad de Guanajuato. Le\'on, Guanajuato, M\'exico.}

\affiliation[b]{Centro de Investigacion en Ciencia Aplicada y Tecnologia Avanzada - Legaria del IPN, M\'exico D.F., M\'exico.}

\affiliation[c]{Departamento de Ingenier\'ia Civil, Divisi\'on de Ingenier\'ia, Universidad de Guanajuato, Guanajuato, M\'exico.}

\affiliation[d]{Instituto de F\'isica y Matem\'aticas, Universidad Michoacana de San Nicol\'as de Hidalgo, Edificio C-3, Ciudad Universitaria, CP. 58040 Morelia, Michoac\'an, M\'exico}

\affiliation[e]{Departamento de Matem\'aticas, Centro Universitario de Ciencias Exactas e Ingenier\'{\i}as (CUCEI), Corregidora 500 S.R., Universidad de Guadalajara, 44420 Guadalajara, Jalisco, M\'exico.}

\emailAdd{avelino@fisica.ugto.mx}
\emailAdd{rigarcias@ipn.mx}
\emailAdd{tamegc72@gmail.com}
\emailAdd{iquiros6403@gmail.com}
\emailAdd{ulises@ifm.umich.mx}

\abstract{ By means of a combined use of the type Ia supernovae {\cred and $H(z)$ data} tests, together with the study of the asymptotic properties in the equivalent phase space -- through the use of the dynamical systems tools -- we demonstrate that the bulk viscous matter-dominated scenario is not a good model to explain the accepted cosmological paradigm, at least, under the parametrization of bulk viscosity considered in this paper. The main objection against such scenarios is the absence of conventional radiation and matter-dominated critical points in the phase space of the model. This entails that radiation and matter dominance are not generic solutions of the cosmological equations, so that these stages can be implemented only by means of unique and very specific initial conditions, i. e., of very unstable particular solutions. Such a behavior is in marked contradiction with the accepted cosmological paradigm which requires of an earlier stage dominated by relativistic species, followed by a period of conventional non-relativistic matter domination, during which the cosmic structure we see was formed. Also, we found that the bulk viscosity is positive just until very late times in the cosmic evolution, around $z < 1$. For earlier epochs it is negative, been in tension with the local second law of thermodynamics.}


\begin{document}

\maketitle


\section{Introduction}


Within the context of early inflation, it has been known since long time ago that an imperfect fluid with bulk viscosity in cosmology can produce an accelerated expansion without the need of a cosmological constant or some other inflationary scalar field \cite{intro1} (although some authors do not agree with this conclusion \cite{intro2}). Further extrapolation of this idea -- used to induce an accelerated expanding Universe without the need of unknown components like dark energy fields -- leads to the possibility that one alternative candidate to explain the present acceleration can be bulk viscous pressure of a imperfect fluid characterizing dark matter \cite{intro3}, in this sense, this class of models represents an unified description of the dark sector in a similar way that the Chaplygin gas model. No matter how attractive it seems, this idea faces some problems, among them, the need to have a satisfactory mechanism for the origin and composition of the bulk viscosity (see, for instance \cite{intro4,intro5}).

From a thermodynamical point of view the bulk viscosity in a physical system is due to its deviations from the local thermodynamic equilibrium \cite{intro5}. In a cosmological setting, the bulk viscosity may arise when the cosmic fluid expands (or contracts) too fast so that the system does not have enough time to restore its local thermodynamic equilibrium and then it arises an effective pressure restoring the system to its thermal equilibrium. When the fluid reaches again the thermal equilibrium then the bulk viscous pressure vanishes \cite{intro5,intro6}. Therefore, in an accelerated expanding Universe, it may be natural to assume the possibility that the expansion process is actually a collection of states out of thermal equilibrium in a small fraction of time giving rise to the existence of a bulk viscosity \cite{ulises}.


Usually the way to test the theoretical (and observational)  viability of a given cosmological model is through using known solutions of the cosmological field equations, or by seeking for new particular solutions that are physically plausible. However, in general, the cosmological field equations are very difficult to solve exactly. Even if a given analytic solution can be found, it will not be unique, but just one in a large set of them. This is not to talk about stability of given solution(s). An alternative way around is to invoke the dynamical systems tools  to extract very useful information about the asymptotic properties of the model instead. In this regard, knowledge of the critical (also equilibrium or singular) points in the phase space -- corresponding to a given cosmological model -- is a very important information since, independent on the initial conditions chosen, the orbits of the corresponding autonomous system of ordinary differential equations (ODE) will always evolve for some time in the
neighborhood of these points. Besides, if the point were an (global) attractor, independent of the initial conditions, the orbits will always be attracted towards it (either into the past or into the future). Going back to the original cosmological model, the existence of the equilibrium points can be correlated with generic cosmological solutions that might really decide the fate and/or the origin of the cosmic evolution.

A phase space for a model which is consistent with  the presently accepted cosmological paradigm should contain critical points associated with: i) a radiation-dominated (relativistic) stage, followed by ii) a matter-dominated (non-relativistic) phase, which is important to allow for the formation of the amount of cosmic structure we see, and iii) an accelerated expanding stage which possibly might last for ever. Perhaps, there should be also a critical point in the phase space associated with an early inflationary period, however, this would require of further refinements of a given cosmological model which is primarily intended to explain the period lasting between decoupling of radiation and baryons and up to the present accelerating phase.

In the present paper we will make a combined use of the dynamical systems tools and of the type Ia supernovae test, to extract as much as possible useful information about the asymptotic properties of a bulk viscous matter-dominated Universe, in order to be able to judge about its theoretical viability to accommodate the accepted cosmological paradigm. In the specific model we shall be investigating the cosmological dynamics is fueled by a conventional baryonic matter component jointly with a fluid with bulk viscosity of the form: $\zeta=\zeta_0+\zeta_1 H+\zeta_2\ddot a/\dot a$ playing the role of dark matter, where $\zeta_0$, $\zeta_1$, and $\zeta_2$ are constants to be determined by the observations and $H=\dot a/a$ is the Hubble parameter. The term $\zeta_0$ takes into account the simplest parametrization for the bulk viscosity: a constant. The term $\zeta_1 H$ characterizes the possibility of a bulk viscosity proportional to the expansion ratio of the Universe, while the third term $\zeta_2\ddot a/\dot a$ takes into consideration the influence acceleration of the expansion might has on the bulk viscosity.

The results of the present study will  convincingly show that the model of bulk viscosity considered here (see also \cite{ulises,ulises1}) is in marked contradiction with the presently accepted cosmological paradigm and, hence, should be ruled out. In fact, it will be shown that there are not any critical points in the phase space of this cosmological model which could be associated with either radiation or matter domination. This, in turn, entails that radiation and matter dominance are not generic solutions of the model. These can be, at most, very particular solutions that can be achieved under specific initial conditions. This is in marked contradiction with the accepted cosmological paradigm in that the latter requires of an earlier stage dominated by relativistic species (specifically a radiation-dominated era), followed by a period of conventional matter domination during which the cosmic
structure we see was formed.

The paper has been organized as  it follows. The relevant details of the model are exposed in section \ref{model}, followed by its observational testing using the supernovae SNe Ia data in section \ref{SNIa}. In the first part of the dynamical systems study in section \ref{d_systems}, for sake of simplicity, we will resort to a single component to embrace radiation, baryons, etc. Then, in subsection \ref{rad}, we consider a more physically involved model where the cosmic dynamics is fueled by radiation, dark matter, and pressureless bulk
viscous matter. In order to draw the phase portraits we use the best estimated values of the free parameters of section \ref{SNIa}. A thorough discussion of the results of our study, based on the combined use of the properties of the model in the equivalent phase space and of the Ia supernovae and $H(z)$ data tests, is given in section \ref{discussion}. It will be evident that, independent of the values of the free parameters, there are not critical points in the phase space that could be correlated with either radiation or matter dominance stages. In the final section \ref{conclusion} brief conclusions are given. An appendix section is included to make the paper self-contained.


\section{Cosmology of bulk viscous matter-dominated Universes.}\label{model}

In this section we analyze a cosmological  model composed by a baryon matter and a bulk viscous components with dust behavior ($w=0$) as dark matter, in a spatially flat universe. The energy-momentum tensor for the pressureless baryon component is given as a perfect fluid as usual $$T^{(m)}_{\mu\nu}=\rho_B\,u_\mu u_\nu,$$ where $\rho_B$ is the energy density of the {\cred baryon} component and $u^\mu$ is the four-velocity vector. The energy-momentum tensor of the bulk viscous component is that of an \textit{imperfect fluid} with  a first-order deviation from the thermodynamic equilibrium.  It can be expressed as \cite{weinberg}: $$T^{(v)}_{\mu\nu}=\rho_v\,u_\mu u_\nu+(g_{\mu\nu}+u_\mu u_\nu)P^*_v,$$ where \be P^*_v\equiv  P_v-\zeta\nabla_{\nu}u^{\nu},\label{pressure-v}\ee $\rho_v$ and $P_v$ are the energy density and pressure of the viscous fluid respectively. The  term $g_{\mu\nu}$ is the metric tensor, and the subscript ``$v$'' stands for ``viscous'' component. The term $\zeta$ is a \emph{bulk viscous} coefficient  that arises in a fluid when it is out of the local thermodynamic equilibrium and which induces a \emph{viscous pressure} $-\zeta\nabla_{\nu}u^{\nu}$ \cite{intro5}. The term $P^*_v$ is an effective pressure composed by the pressure $P_v$ of fluid plus the bulk viscous pressure. It was initially proposed in reference \cite{eckart} for relativistic dissipative processes in thermodynamic systems out of local equilibrium, and later on the authors of Ref. \cite{landau} developed an equivalent formulation.

Here we shall explore the cosmological implications of such a model by assuming that the Universe is filled with bulk viscous fluid only. For Friedmann-Robertson-Walker (FRW) spacetimes with flat spatial sections we have $$ds^2=-dt^2+a^2(t)(dr^2+r^2d\Omega^2),$$ where $a(t)$ is the scale factor. The conservation equations for the pressureless baryon matter and the viscous component can be written respectively as
\bea &&\dot{\rho}_B+3H\rho_B=0,\nonumber\\
&&\dot{\rho}_v+3H(\rho_v+P^*_v)=0,\label{cons_eq_vf}\eea where $H \equiv \dot{a}/a$ is the Hubble parameter and, as usual, the over-dot stands for time derivative. The solution of the first equation in (\ref{cons_eq_vf}) for the matter component is $\rho_B=\rho_{B0}a^{-3}$, where $\rho_{B0}$ is the present value of the matter energy density. On the other hand, the bulk viscous pressure $-\zeta\nabla_{\nu}u^{\nu}$ can be written as $-3\zeta H$. So, assuming the general equation of state for the viscous component as $P_v=w\rho_v$, the second conservation equation in (\ref{cons_eq_vf}) for the viscous fluid becomes
\be \dot{\rho}_v+3H(1+w)\rho_v-9\zeta H^2=0.\label{cons_eq_vf1}\ee

The Einstein's equations for the model are
\bea &&H^2=\frac{8\pi G}{3}\left(\rho_B+\rho_v\right),\;\frac{\ddot{a}}{a}=-\frac{4\pi G}{3} \Bigl[\rho_B+(1+3w)\rho_v-9\zeta
H\Bigr].\label{feqs}\eea

We assume the parametrization for the bulk viscosity $\zeta$ of  the viscous component $\rho_v$ in the form of the following expansion:
\be \zeta=\zeta_0+\zeta_1 H+\zeta_2\left(\frac{\ddot{a}}{\dot{a}}\right),\label{par_z}\ee where $\zeta_0$, $\zeta_1$ and $\zeta_2$ are constants to be determined by the observations. The term $(\ddot{a}/\dot{a})$ in (\ref{par_z}) can be written as $(\ddot{a}/\dot{a})= (\ddot{a}/a)/H$.  So, using second equation in (\ref{feqs}), it can be written as
\be \frac{\ddot{a}}{\dot{a}}= -\frac{4\pi G}{3H}\Bigl[\rho_B+(1+3w)\rho_v - 9\zeta H\Bigr].\label{EquivalentFriedmann2Eq}\ee
{\cred We can see that if the magnitude of the term $9\zeta H$ is greater than ``$\rho_B+(1+3w)\rho_v$'' then it is induced an acceleration in the expansion of the Universe, i.e., $\ddot{a}>0$}.

For simplicity, we define the dimensionless bulk viscous coefficients as
\bea &&\tilde{\zeta}\equiv\left(\frac{8\pi G}{3H_0}\right)\zeta, \qquad \;\tilde{\zeta}_0\equiv\left(\frac{24\pi G}{H_0}\right)\zeta_0 \label{DimensionlessViscosityDefinitions} \\
&&\tilde{\zeta}_1\equiv\left(24\pi G\right)\zeta_1, \qquad \;\tilde{\zeta}_2\equiv\left(\frac{4\pi G}{3}\right)\zeta_2. \nonumber\eea
The dimensionless Friedmann constraint -- first equation in (\ref{feqs}) -- can be written as
\be E^2\equiv\left(\frac{H}{H_0}\right)^2=\Omega_{B0}a^{-3}+\tilde{\Omega}_v,\label{Friedmann1st-Dimensionless}\ee where $\Omega_{B0}\equiv\rho_{B0}/\rho_{\rm crit}^0$ and $\tilde{\Omega}_v \equiv \rho_v/\rho_{\rm crit}^0$. The evolution of the viscous density $\tilde{\Omega}_v$ is given by the numerical solution of the ordinary differential equation (\ref{ODE2}). See Appendix \ref{SectionAppendixOmega} for details.


\begin{figure}[t!]
\begin{center}
\includegraphics[width=8cm,height=7cm]{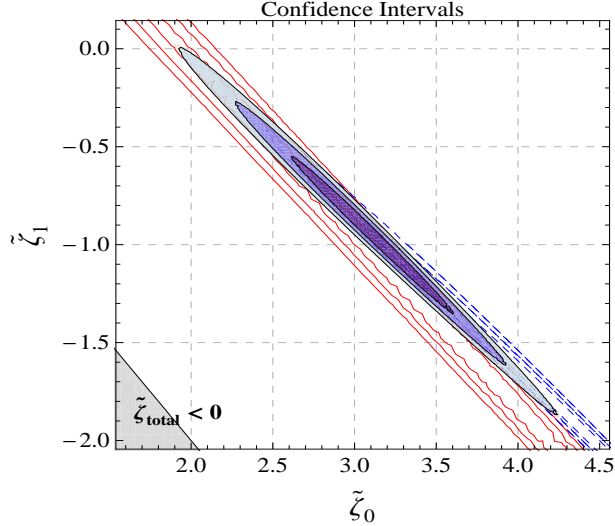}
\caption{Confidence intervals for the dimensionless viscous coefficients $(\tilde{\zeta}_0,\tilde{\zeta}_1)$ of a bulk viscosity parametrization of the form $\zeta=\zeta_0 + \zeta_1 H$ (we have set $\tilde{\zeta}_2=0$). We find as best estimates $(\tilde{\zeta}_0=3.12^{+0.35}_{-0.34}$, $\tilde{\zeta}_1=-0.96 \pm 0.3)$, see table \ref{best-estimated}. The solid red, dashed and filled contour plots correspond to the use of SNe, $H(z)$ and the joint SNe + $H(z)$ data sets respectively. The gray shaded area at the bottom left indicates the \textit{forbidden} region where the total bulk viscosity {\cred $\tilde{\zeta}_{\rm }$} given by eq.(\ref{ViscosityEq1-Dimensionless}) is \textit{negative} at the present-day ($z=0$). The confidence intervals shown correspond to $68.3\% $, $95.4\%$ and $99.73\%$ of probability. The Hubble constant $H_0$ was marginalized assuming a \textit{constant prior} probability distribution for $H_0$. It was assumed the values $\tilde{\Omega}_v(z=0)=0.96$, $\Omega_{B0}=0.04$ and $w=0$.}
\label{plotJustCIZ01}
\end{center}
\end{figure}


\begin{figure}[t!]
\begin{center}
\includegraphics[width=8cm,height=7cm]{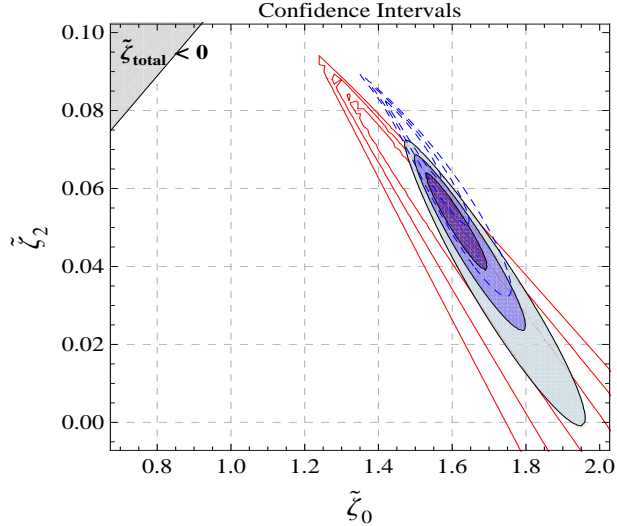}
\caption{Confidence intervals for $(\tilde{\zeta}_0,\tilde{\zeta}_2)$ of a bulk viscosity parametrization of the form $\zeta=\zeta_0+\zeta_2 (\ddot{a}/\dot{a})$ (we set $\tilde{\zeta}_1=0$). We find as best estimates $(\tilde{\zeta}_0=1.59^{+0.06}_{-0.05}$, $\tilde{\zeta}_2=0.05^{+0.007}_{-0.01})$, see table \ref{best-estimated}. The solid red, dashed and filled contour plots correspond to the use of SNe, $H(z)$ and the joint SNe + $H(z)$ data sets respectively. The gray shaded area at the top left indicates the \textit{forbidden} region
where the total bulk viscosity {\cred $\tilde{\zeta}_{\rm }$ } given by eq.(\ref{ViscosityEq1-Dimensionless}) is \textit{negative} at the present-day ($z=0$). The confidence intervals shown correspond to $68.3\% $, $95.4\%$ and $99.73\%$ of probability.}\label{plotJustCIZ02}
\end{center}
\end{figure}


\begin{figure}[t!]
\begin{center}
\includegraphics[width=8cm,height=7cm]{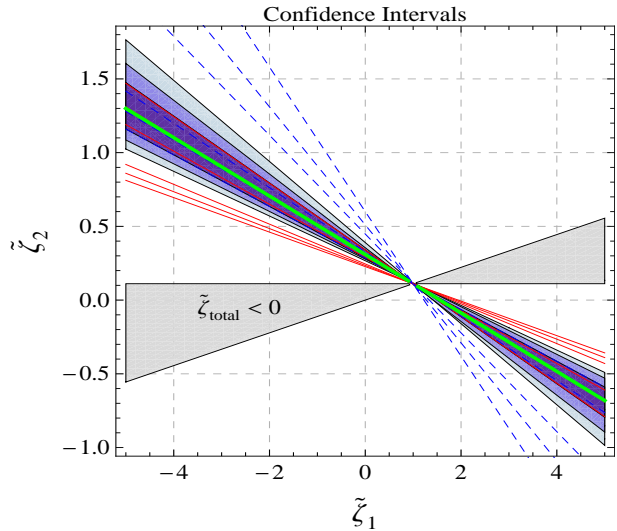}
\caption{Confidence intervals for $(\tilde{\zeta}_1,\tilde{\zeta}_2)$ of a bulk viscosity parametrization of the form $\zeta=\zeta_1 H+\zeta_2 (\ddot{a}/\dot{a})$  (we set $\tilde{\zeta}_0=0$). The solid red, dashed and filled contour plots correspond to the use of SNe, $H(z)$ and the joint SNe + $H(z)$ data sets respectively. The confidence intervals shown correspond to $68.3\% $, $95.4\%$ and $99.73\%$ of probability. The best estimated values for $(\tilde{\zeta}_1,\tilde{\zeta}_2)$ are \textit{all} those points that lie on the line equation $\tilde{\zeta}_2=m \tilde{\zeta}_1+b$, where $m = -0.19808$, $b=0.30918$ with $\chi^2_{\rm min}= 639.131$ (green line) when using the combined SNe + $H(z)$ data sets,
except at the singular point $(\tilde{\zeta}_1 = 1,\tilde{\zeta}_2=1/9)$ that corresponds to the vertex visible in the figure. The gray shaded area indicates the \textit{forbidden} region where the total bulk viscosity {\cred $\tilde{\zeta}_{\rm }$ } given by eq.(\ref{ViscosityEq1-Dimensionless}) is \textit{negative} at the present day ($z=0$).}\label{plotJustCIZ12}
\end{center}
\end{figure}



\begin{table}\centering
\begin{tabular}{ l  l | c | c c }
\multicolumn{5}{c}{\textbf{Viscous model}}\\
\hline \hline
\multicolumn{2}{|c|}{Best estimates} &  Assumption & $\chi^2_{{\rm
min}}$ & $\chi^2_{{\rm d.o.f.}}$ \\
\hline $\tilde{\zeta}_0=3.12^{+0.35}_{-0.34}$  &  $\tilde{\zeta}_1 = -0.96 \pm 0.3$ & $\tilde{\zeta}_2 = 0$  & 573.34 & 0.97 \\
$\tilde{\zeta}_0 = 1.59^{+0.06}_{-0.05} $  &  $\tilde{\zeta}_2=0.05^{+0.007}_{-0.01}$ & $\tilde{\zeta}_1 = 0 $  & 573.34 & 0.97 \\
\hline \hline\end{tabular}
\caption{Best estimated values of the dimensionless viscous coefficients  $(\tilde{\zeta}_0, \tilde{\zeta}_1, \tilde{\zeta}_2)$ for the dark component with bulk viscosity parametrized as $\zeta=\zeta_0+\zeta_1 H+\zeta_2 (\ddot{a}/\dot{a})$. It was assumed $w=0$ and $\tilde{\Omega}_v(z=0)=0.96$ for the dark component, and $\Omega_{B0}=0.04$ for the baryon matter. The best estimates were computed using the SNe + $H(z)$ data sets. The first two columns show the best estimated values for pairs of viscous coefficients $(\tilde{\zeta}_i,\tilde{\zeta}_j)$, where the remaining viscous coefficient $\tilde{\zeta}_k$ in each case, is set to zero (third column). The fourth and fifth columns show the minimum of the $\chi^2$ function and its corresponding ``$\chi^2$ function by \textit{degrees of freedom}'': $\chi^2_{\rm d.o.f.}\equiv\chi^2_{\rm min}/(n-p)$, where $n$ is the number of data, and $p$ is the number of free parameters estimated. The errors in the estimations are given to 1$\sigma$. The best estimated values for $(\tilde{\zeta}_1,\tilde{\zeta}_2)$, with $\tilde{\zeta}_0=0$, correspond to \textit{all} those points which lie on the line $\tilde{\zeta}_2=m\tilde{\zeta}_1+b$ ($m=-0.19808$, $b=0.30918$), with $\chi^2_{\rm min}=639.131$ ($\chi^2_{\rm d.o.f.}=1.08$). Figures \ref{plotJustCIZ01} to \ref{plotJustCIZ12} show the confidence intervals. The Hubble constant $H_0$ was marginalized assuming a \textit{constant} prior distribution.}\label{best-estimated}
\end{table}


\section{Cosmological probes}\label{SectionObservationalConstraints}

We test the viability of the model and constrain its free parameters $(\tilde{\zeta}_0,\tilde{\zeta}_1,\tilde{\zeta}_2)$ using the type Ia Supernovae (SNe Ia) observations and the Hubble parameter $H(z)$ measured at different redshifts. We compute the \textit{best estimated values} for pairs of $(\tilde{\zeta}_0,\tilde{\zeta}_1,\tilde{\zeta}_2)$, the \emph{goodness-of-fit} of the model  to the data and the confidence intervals by a $\chi^2$ function minimization, to constrain their possible values with levels of statistical confidence which are shown in figures \ref{plotJustCIZ01}--\ref{plotJustCIZ12}.


\begin{figure}
\begin{center}
\includegraphics[width=7cm, height=4.5cm]{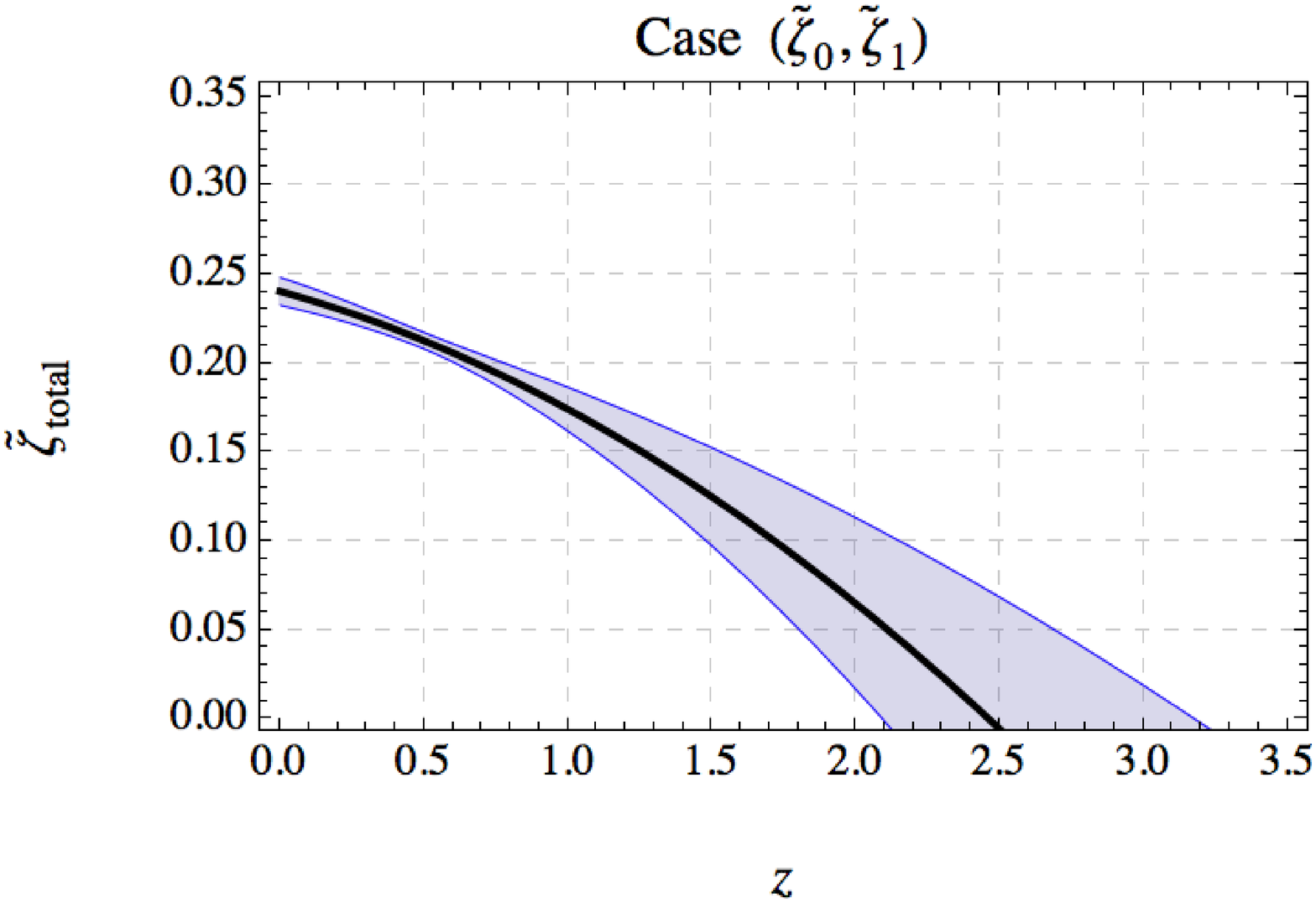}
\includegraphics[width=7cm, height=4.5cm]{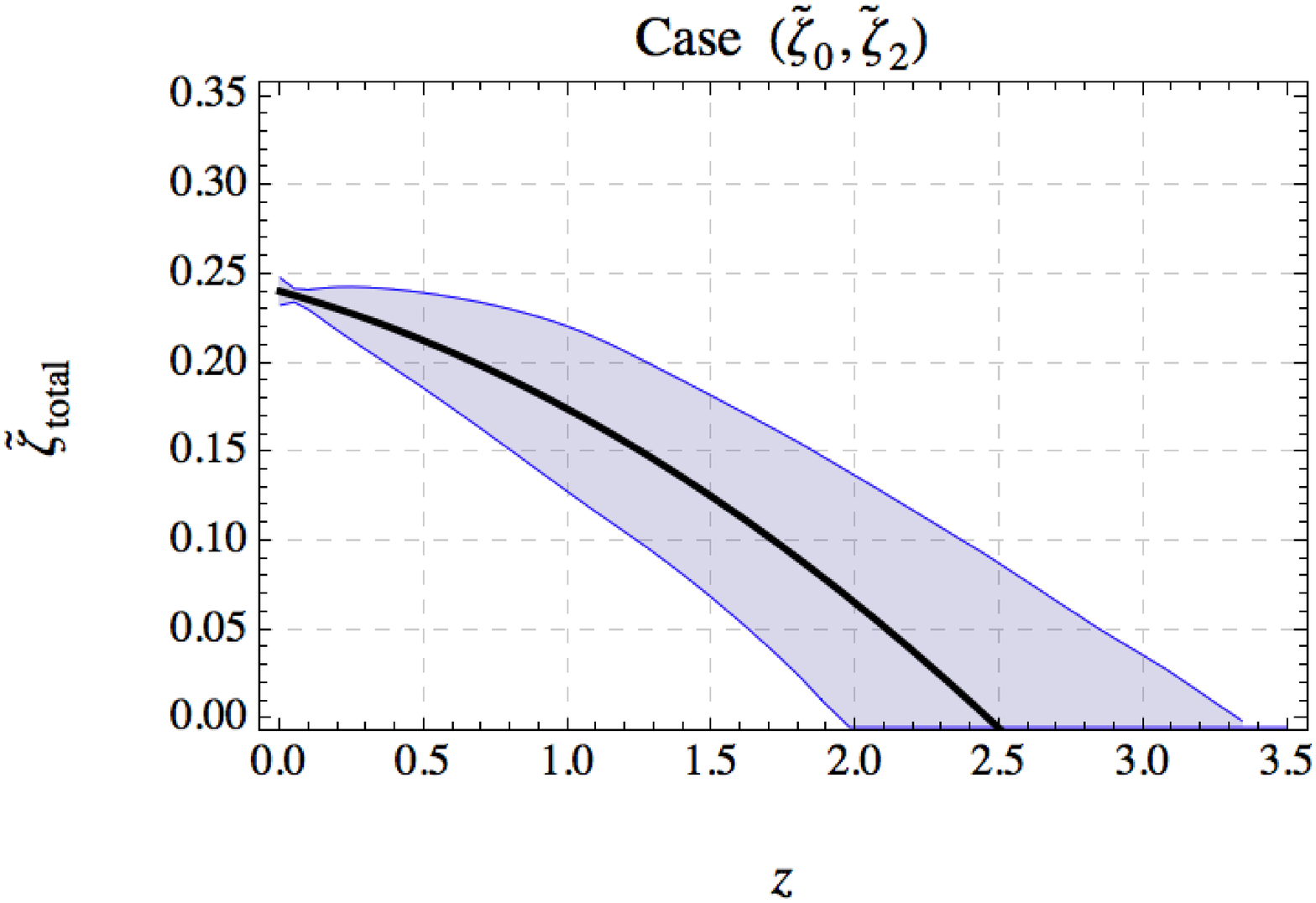}
\caption{{\cred Evolution of the \textit{total} bulk viscosity $\tilde{\zeta}_{\rm }$ with respect to the redshift $z$. The left and right panels correspond to the cases $(\tilde{\zeta}_0, \tilde{\zeta}_1)$ and $(\tilde{\zeta}_0, \tilde{\zeta}_2)$ respectively. The central thick lines come from the evaluation of the equation (\ref{ViscosityEq1-Dimensionless}) at the best estimated values for $(\tilde{\zeta}_0, \tilde{\zeta}_1)$ and $(\tilde{\zeta}_0, \tilde{\zeta}_2)$ respectively (see table \ref{best-estimated}). We notice that total bulk viscosity $\tilde{\zeta}$ arises with positive values just at late times, at around a redshift $z=2.5$. It is not shown the negative values of $\tilde{\zeta}_{\rm }$ because they are forbidden by the local second law of thermodynamics (see section \ref{LSLT}). The error bands are given at 68.3\% ($1\sigma$) of confidence level (see Appendix \ref{SectionErrorPropagation} for details).}}\label{PlotZetaTotal}
\end{center}
\end{figure}



\begin{figure}
\begin{center}
\includegraphics[width=7cm, height=4.5cm]{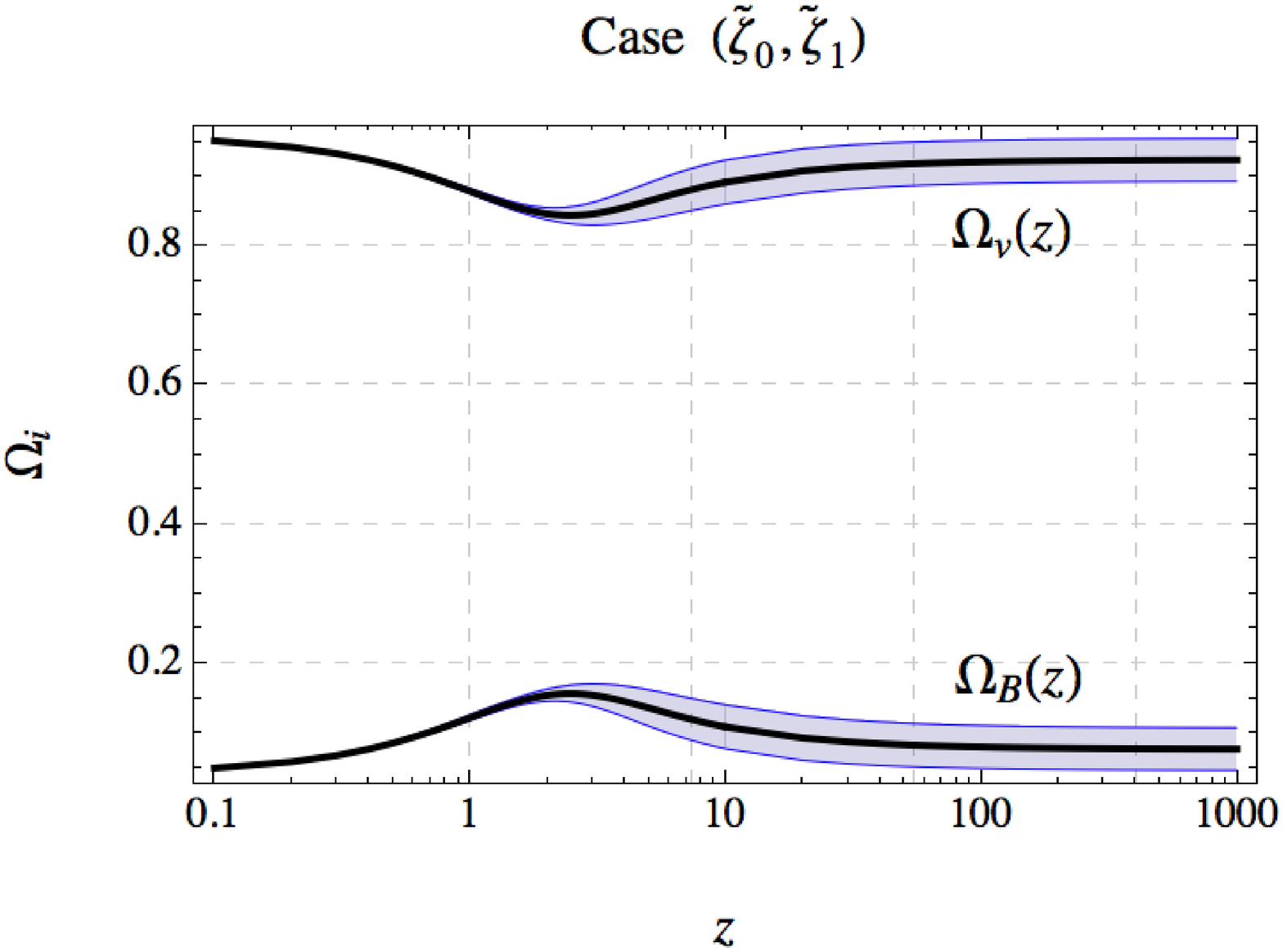}
\includegraphics[width=7cm, height=4.5cm]{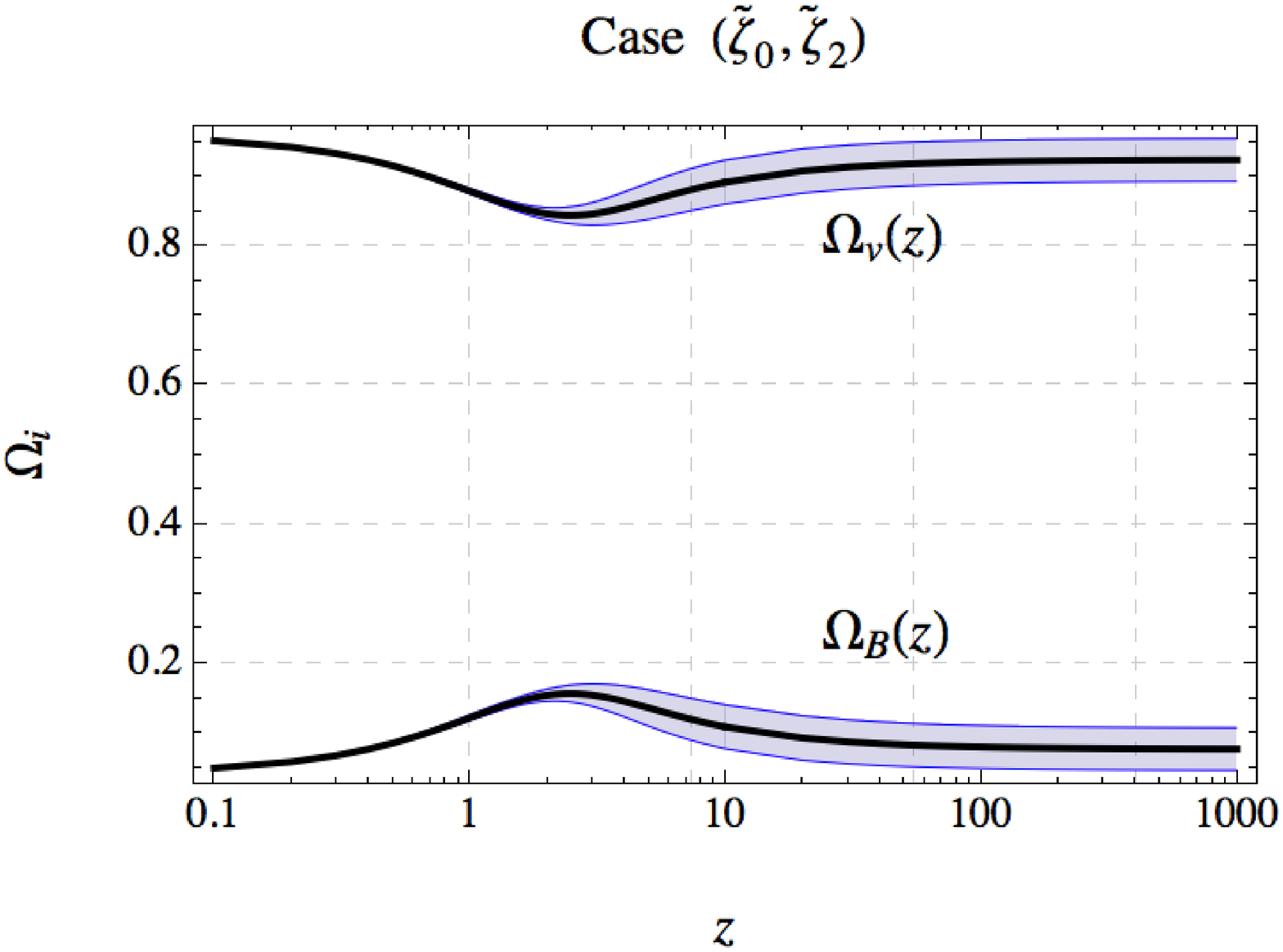}
\caption{{\cred Evolution of the dimensionless density parameters of the viscous dark matter component $\Omega_{v}(z)$ [upper curve] and  the baryon component $\Omega_{B}(z)$ [lower curve] with respect to the redshift $z$, for the cases $(\tilde{\zeta}_0, \tilde{\zeta}_1)$ and $(\tilde{\zeta}_0, \tilde{\zeta}_2)$. The thick black lines come from the evaluation of the equation (\ref{Eq-bothOmegasZ}) at the best estimated values for $(\tilde{\zeta}_0, \tilde{\zeta}_1)$ and $(\tilde{\zeta}_0, \tilde{\zeta}_2)$ (see table \ref{best-estimated}). The error bands are given at 68.3\% ($1\sigma$) of confidence level (see Appendix \ref{SectionErrorPropagation} for details). We see that $\Omega_{v}(z)$ is always the dominant component with respect to $\Omega_{m}(z)$ at late times, so that . It is seen a bouncing in the evolution of the densities at a redshift of around $z=2.5$, that corresponds when the total bulk viscosity $\tilde{\zeta}$ has a transition from negative to positive values as $z \rightarrow 0$ (see figure \ref{PlotZetaTotal}). }}\label{PlotOmegaZ}
\end{center}
\end{figure}



\begin{figure}[t!]
\begin{center}
\includegraphics[width=7cm, height=4.5cm]{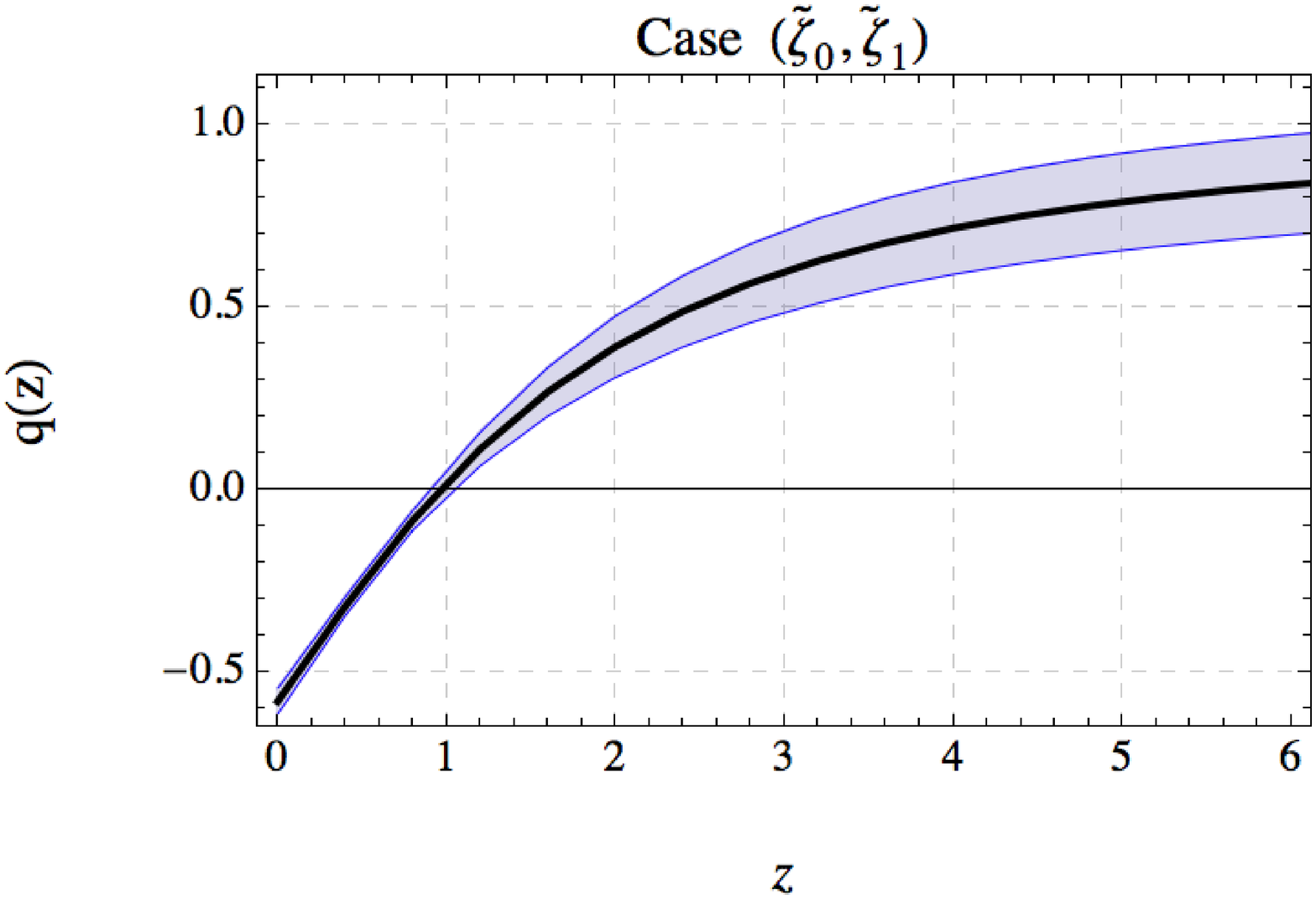}
\includegraphics[width=7cm, height=4.5cm]{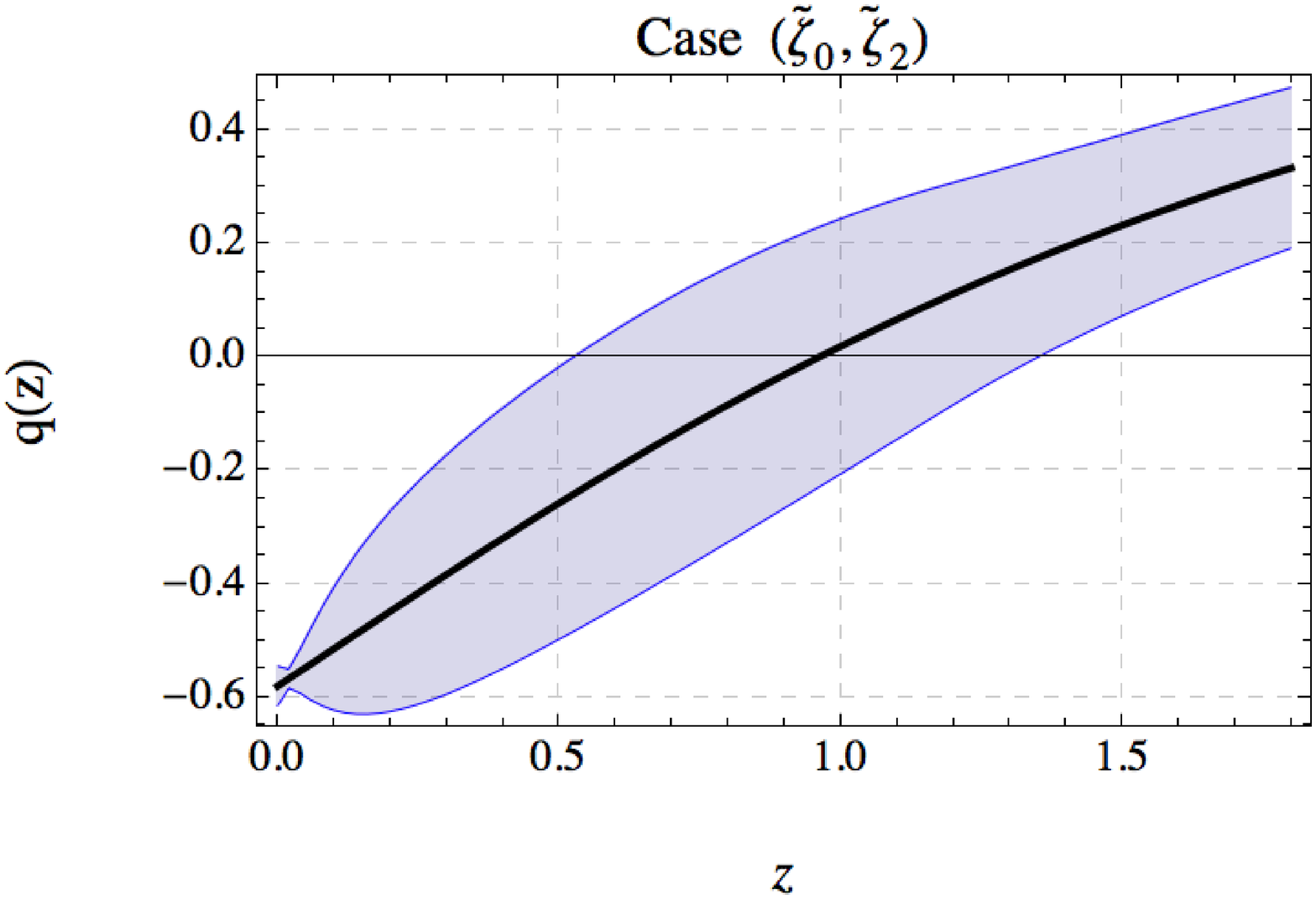}
\caption{   {\cred Evolution of the deceleration parameter $q(z)$ with respect to the redshift $z$ for the cases $(\tilde{\zeta}_0, \tilde{\zeta}_1)$ and $(\tilde{\zeta}_0, \tilde{\zeta}_2)$. The thick black lines come from the evaluation of the equation (\ref{EqqDecelerationZ}) at the best estimated values for $(\tilde{\zeta}_0, \tilde{\zeta}_1)$ and $(\tilde{\zeta}_0, \tilde{\zeta}_2)$ (see table \ref{best-estimated}). We notice that both cases predict an accelerating expanding Universe at late times, with a transition from deceleration to acceleration at about $z \simeq 1$ The error bands are given at 68.3\% ($1\sigma$) of confidence level (see Appendix \ref{SectionErrorPropagation}).}}\label{PlotqDecelerationZ}
\end{center}
\end{figure}


\subsection{Type Ia Supernovae}\label{SNIa}

We use the ``Union2.1'' SNe Ia data set (2012) from ``The Supernova Cosmology Project'' (SCP) composed by 580 type Ia supernovae \cite{SuzukiUnion2.1-2012}. The luminosity distance $d_L$ in a flat FRW cosmology, is computed through $$d_L(z,\tilde{\zeta}_0,\tilde{\zeta}_1,\tilde{\zeta}_2,H_0)=\frac{c (1+z)}{{\cred H_0}} \int_0^z\frac{dz'}{E(z',\tilde{\zeta}_0,\tilde{\zeta}_1,\tilde{\zeta}_2)},$$ where $E(z, \tilde{\zeta}_0,\tilde{\zeta}_1, \tilde{\zeta}_2)$ is given by the expression (\ref{Friedmann1st-Dimensionless}) and $c$ is the speed of light
given in units of km/sec. The theoretical distance moduli for the $k$-th supernova with redshift $z_k$ is defined as

\begin{equation}\label{distanceModuli}
\mu^{{\rm t}}(z_k,\tilde{\zeta}_0,\tilde{\zeta}_1,\tilde{\zeta}_2,H_0)\equiv m - M = 5\log_{10}\left[\frac{d_L(z_k,\tilde{\zeta}_0,\tilde{\zeta}_1,\tilde{\zeta}_2, H_0)}{1 \, {\rm Mpc}}\right]+25,\end{equation}

\noindent where $m$ and $M$ are the apparent and absolute magnitudes of the SNe Ia respectively, and the superscript `t' stands for \textit{``theoretical''}. We construct the statistical $\chi^2$ function as
\be \chi^2_{\rm SNe}(\tilde{\zeta}_0,\tilde{\zeta}_1,\tilde{\zeta}_2)\equiv\sum_{k=1}^n\frac{\left[\mu^{{\rm t}}(z_k,\tilde{\zeta}_0,\tilde{\zeta}_1,\tilde{\zeta}_2)-\mu_k\right]^2}{\sigma_k^2},\label{ChiSquareDefinition}\ee where $\mu_k$ is the \emph{observational} distance moduli for the $k$-th supernova, $\sigma_k^2$ is the variance of the measurement and
$n$ is the amount of supernova in the data set ($n=580$). The results are shown in table \ref{best-estimated} and the confidence intervals for the pairs $\tilde{\zeta}_0$ vs $\tilde{\zeta}_1$,$\tilde{\zeta}_0$ vs $\tilde{\zeta}_2$ and $\tilde{\zeta}_1$ vs $\tilde{\zeta}_2$ are shown in figures \ref{plotJustCIZ01} to \ref{plotJustCIZ12} respectively. The Hubble constant $H_0$ is marginalized assuming a \textit{constant prior} distribution (see appendix A of \cite{ulises1}).

\subsection{Hubble expansion rate}

For the Hubble parameter $H(z)$  measured at different redshifts, we use the 12 data listed in table 2 of Busca et al. (2012) \cite{Busca}, where 11 data come from references \cite{Blake1}--\cite{RiessHzToday}. The value $H_0=70$ km s$^{-1}$ Mpc$^{-1}$, is assumed for the data of Blake et al. (2011) \cite{Blake1} as Busca et al. suggest. The $\chi^2$ function is defined as
\be \chi^2_{\rm H}(\tilde{\zeta}_0,\tilde{\zeta}_1,\tilde{\zeta}_2)=\sum_i^{12} \left(\frac{H(z_i,\tilde{\zeta}_0,\tilde{\zeta}_1,\tilde{\zeta}_2)-H_i^{\rm obs}}{\sigma_{H i}}\right)^2,\label{Chi2Hz}\ee where $H(z_i, \tilde{\zeta}_0,\tilde{\zeta}_1,\tilde{\zeta}_2)$ and $H_i^{\rm obs}$ are the theoretical and observed values respectively and $\sigma_{H i}$ the standard deviation of each $H_i^{\rm obs}$ entry.


The \textit{total} $\chi^2_{\rm t}$ function which  combines the SNe and $H(z)$ data sets together, is chosen in the following way:
\begin{equation}\label{ChiSqrTotal}
\chi^2_{\rm t}=\chi^2_{\rm SNe}+\chi^2_{\rm H},\end{equation} where $\chi^2_{\rm SNe}$ and $\chi^2_{\rm H}$ are given by expressions (\ref{ChiSquareDefinition}) and (\ref{Chi2Hz}) respectively. The function $\chi^2_{\rm t}$ is then numerically minimized in order to compute the ``\textit{best estimates}'' for pairs of the viscous coefficients: $(\tilde{\zeta}_0,\tilde{\zeta}_1)$, $(\tilde{\zeta}_0,\tilde{\zeta}_2)$, and $(\tilde{\zeta}_1,\tilde{\zeta}_2)$, where the remaining viscous coefficient in each case is assumed to vanish. The minimum
value of the $\chi^2$ function gives the best estimated values of the pairs $(\tilde{\zeta}_0,\tilde{\zeta}_1)$, $(\tilde{\zeta}_0,\tilde{\zeta}_2)$, and $(\tilde{\zeta}_1,\tilde{\zeta}_2)$, and measures the goodness-of-fit of the model to data.

The definition of ``$\chi^2$ function by \textit{degrees of freedom}'': $\chi^2_{\rm d.o.f.}\equiv\chi^2_{\rm min}/(n-p)$, where $n$ is the number of \textit{total} combined data used, and $p$ is the number of free parameters estimated, is also used in our computations.

\subsection{Local Second Law of Thermodynamics}\label{LSLT}

The \textit{local} entropy production for a fluid on a FRW spacetime is expressed as \cite{weinberg}

\be T\,\nabla_{\nu} s^{\nu}=\zeta(\nabla_{\nu} u^{\nu})^2=9H^2\zeta,\label{entropy_definition}\ee where $T$ is the temperature of the fluid, $\nabla_{\nu} s^{\nu}$ is the rate of entropy production in a unit volume, and $\zeta$ is the \textit{total} bulk viscosity. The second law of the thermodynamics can be stated as $T\nabla_{\nu} s^{\nu}\geq 0$. Hence, since the Hubble parameter $H$ is positive for an expanding Universe, Eq. (\ref{entropy_definition}) implies that $\tilde{\zeta}\geq 0$, where $\tilde{\zeta}$ is given by the expression (\ref{ViscosityEq1-Dimensionless}). This inequality is an additional constraint in the possible values for the \textit{total} dimensionless viscous parameter in our model.


\section{Dynamical Systems}\label{d_systems}

The dynamical systems tools offer a powerful mean to extract relevant information out of the given cosmological model by investigating the equivalent phase space. Critical points in the phase space: past/future attractors, saddle points, etc., can be correlated with generic solutions of the cosmological field equations (see \ref{app-d-systems}). In order to be able to apply these tools one has to follow the steps enumerated here: i) to identify the phase space variables that allow writing the system of cosmological equations in the form of an autonomous system of ODE,\footnote{There might be several different possible choices, however, not all of them allow for the minimum possible dimensionality of the phase space.} ii) with the help of the chosen phase space variables to build an autonomous system of ODE out of the original system of cosmological equations, and an usually forgotten or unappreciated step, iii) to identify the phase space spanned by the chosen variables that is relevant to the cosmological model under study.

Our goal in this section is to write the cosmological equations of the model (\ref{cons_eq_vf}-\ref{feqs}):\footnote{Here we use units where $8\pi G=1$.}

\bea &&\dot \rho_B + 3H\rho_B=0, \qquad  \dot\rho_v + 3H(1+\omega)\rho_v-9\zeta H^2=0,\nonumber\\
&&3 H^2=\rho_v+\rho_B, \qquad 6\dot H+6 H^2=-\rho_B-(1+3\omega)\rho_v+9\zeta H,\label{feqs_1}\eea in the form of an autonomous system of ODE. To this end we have to choose appropriate phase space variables. In the present case our starting phase space variable is the dimensionless energy density parameter of the viscous fluid:

\be x\equiv\Omega_v=\frac{\rho_v}{3H^2},\qquad 0\leq x\leq1.\label{var_x}\ee

In terms of this variable the Friedmann constraint (third equation in (\ref{feqs_1})) can be written as $\Omega_B=1-x$, where we use the standard definition of the dimensionless energy density parameter of the $i$-th matter component, $\Omega_i\equiv\rho_i/3H^2$. Also, one can write the following autonomous ODE:

\be x'=-3(1+\omega)x-2x\frac{H'}{H}+3\frac{\zeta}{H},\label{odex}\ee or, since

\be 2\frac{H'}{H}=-3(1+\omega x)+3\frac{\zeta}{H},\label{h'/h}\ee the former equation can be written in more compact form:

\be x'=3(x-1)\left(\omega x-\frac{\zeta}{H}\right).\label{odex_g}\ee In the above equations the tilde accounts for derivative with respect to the parameter, $\tau=\ln a$.

As already mentioned, here we shall investigate a viscous coefficient of the form given in equation (\ref{par_z}): $\zeta=\zeta_0+\zeta_1 H+\zeta_2 \ddot a/\dot a$, or, since $\ddot a/a=\dot H+H^2$, then the viscous parameter will obey the following equation:

\bea \frac{\zeta}{H}=\frac{\zeta_0}{H}+\zeta_1+\zeta_2+\frac{H'}{H}\zeta_2=\frac{2\zeta_0/H+2\zeta_1-(1+3\omega x)\zeta_2}{2-3\zeta_2},\label{zeta/h}\eea where, in the last row, we have taken into consideration equation (\ref{h'/h}). If we substitute back Eq.(\ref{zeta/h}) into (\ref{odex_g}), we obtain the following master ODE:

\be x'=\frac{6(x-1)}{2-3\zeta_2}\left(\omega x-\frac{\zeta_0}{H}-\zeta_1+\frac{\zeta_2}{2}\right).\label{main}\ee

Several cosmological parameters, such as the deceleration parameter $q=-1-H'/H$, and the equation of state (EoS) effective parameter $\omega_{eff}=-1-2H'/3H$, can also be rewritten in terms of the variable $x$. In fact, if take into account equations (\ref{h'/h}) and (\ref{zeta/h}) one obtains:

\bea q=\frac{1+3\omega x-3\zeta_0/H-3\zeta_1}{2-3\zeta_2},\qquad \omega_{eff}=\frac{2\omega x-2\zeta_0/H-2\zeta_1+\zeta_2}{2-3\zeta_2},\label{parameters}\eea respectively.

Depending on the particular case of (\ref{zeta/h}) under consideration one would need yet another phase space variable which would be related with the viscous coefficient $\zeta_0$ (see below). In what follows we shall split the dynamical systems study into two different cases: i) when the viscous EoS parameter $\omega\neq 0$, and ii) when the viscous fluid is dust: $\omega=0$.

\subsection{Viscous EoS $\omega\neq 0$}

\subsubsection{Case with $\zeta_0=0$.}\label{zeta0=0}

If we set $\zeta_0=0$ in Eq.(\ref{main}), we obtain the following autonomous ODE for this particular case:

\be x'=\frac{6\omega (x-1)}{2-3\zeta_2}\left(x-\frac{2\zeta_1-\zeta_2}{2\omega}\right).\label{odex1}\ee The phase space is the segment, $\Psi=\{x|0\leq x\leq1\}$.

Two equilibrium/critical points are found:

\begin{enumerate}

\item The first one, $$P_v:\left(x=\frac{\rho_v}{3H^2}=1\right),$$ corresponds to the viscous matter-dominated solution. The deceleration and EoS effective parameters, in this case, are given by: $$q=\frac{1+3(\omega-\zeta_1)}{2-3\zeta_2},\qquad \omega_{eff}=\frac{2(\omega-\zeta_1)+\zeta_2}{2-3\zeta_2}.$$ The solution corresponds to accelerated expansion whenever, either $\zeta_1>(1+3\omega)/3$, $\zeta_2<2/3$, or, $\zeta_1<(1+3\omega)/3$, $\zeta_2>2/3$. Otherwise it will correspond to decelerated expansion instead.

If consider small $\tau$-dependent perturbation $\epsilon=\epsilon(\tau)$ around this critical point: $x\rightarrow 1+\epsilon(\tau)$, up to ${\cal O}(\epsilon^2)$, the perturbation will obey the following linearized ODE: $\epsilon'(\tau)=\lambda \epsilon(\tau)$, which can be readily integrated, $$\epsilon(\tau)=\epsilon_0\;e^{\lambda\tau}, \qquad \lambda=3\frac{\zeta_2-2\zeta_1+2\omega}{2-3\zeta_2},$$ where $\epsilon_0$ is an integration constant. The solution is stable or, in other words, it is a future attractor in the phase segment if $\lambda<0$, i. e., if either $2\zeta_1-\zeta_2>2\omega,\;\zeta_2<2/3$, or $2\zeta_1-\zeta_2<2\omega,\;\zeta_2>2/3$, which coincide with the regions in the space of parameters $(\zeta_1,\zeta_2,\omega)$ where the point $P_v$ is correlated with inflationary expansion ($q<0$). Otherwise, if $2\zeta_1-\zeta_2<2\omega,\;\zeta_2<2/3$, or $2\zeta_1-\zeta_2>2\omega,\;\zeta_2>2/3$, the viscous matter-dominated solution is unstable (it is a past attractor), while the expansion occurs at a decelerated pace. Hence, either the viscous matter-dominated critical point $P_v$ ($3H^2=\rho_v$), is the future inflationary attractor/end-point of any phase space orbit, or, alternatively, it is the past attractor/source point in the phase space, which is associated with decelerated expansion.

\item The second critical point, $$P_{B/v}:\left(x=\frac{2\zeta_1-\zeta_2}{2\omega}\right)\;\Rightarrow\;3H^2=\frac{2\omega\rho_v}{2\zeta_1-\zeta_2},$$ exists whenever\footnote{We shall be assuming that $\omega$ is a non-negative quantity, which covers the most interesting physical situations.} $0<2\zeta_1-\zeta_2\leq 2\omega$, and corresponds to matter/viscous matter-scaling solution: $$\frac{\Omega_B}{\Omega_v}=\frac{2\omega-2\zeta_1+\zeta_2}{2\zeta_1-\zeta_2}.$$ The deceleration and the EoS parameters for this critical point are: $q=1/2$ and $\omega_{eff}=0$ respectively. Small $\tau$-dependent perturbations around this critical point will obey, $\epsilon'=\lambda\epsilon$, or after integration, $$\epsilon(\tau)=\epsilon_0\;e^{\lambda\tau}, \qquad \lambda=\frac{2\zeta_1-\zeta_2-2\omega}{2-3\zeta_2}.$$ Hence the scaling critical point is stable, i. e., it is the future attractor in the phase space, if $\zeta_2<2/3$. Otherwise, if $\zeta_2>2/3$, this equilibrium point is the past attractor. In consequence, when the scaling equilibrium point exists, either it is the future attractor while the first equilibrium point ($x=1$) is the past attractor, or vice versa.

\end{enumerate}


\begin{table}\centering
\begin{tabular}{| c | c | c | c |}
\multicolumn{4}{c}{\textbf{Critical points ($\omega\neq 0$, $\zeta_0=0$)}}\\
\hline \hline
$P_i$ &  $x$ & Existence & Stability\\
\hline\hline
$P_v$ & $1$ & Always& Stable if $2\zeta_1-\zeta_2>2\omega,\;\zeta_2<2/3$,\\
 & & & or if $2\zeta_1-\zeta_2>2\omega,\;\zeta_2<2/3$.\\
& & & Unstable if $2\zeta_1-\zeta_2<2\omega,\;\zeta_2<2/3$,\\
& & & or if $2\zeta_1-\zeta_2>2\omega,\;\zeta_2>2/3$.\\
\hline
 $P_{B/v}$& $\frac{2\zeta_1-\zeta_2}{2\omega}$ & $6<2\zeta_1-\zeta_2\leq 2\omega$&  Stable if $\zeta_2<2/3$,\\
& & & Unstable if $\zeta_2>2/3$.\\
\hline\hline
\end{tabular}\caption{Existence and stability of the critical points $P_i$ for the particular case when $\omega\neq 0$, $\zeta_0=0$.}\label{tab1}
\end{table}


The properties of these equilibrium points are summarized in table \ref{tab1}. In this particular case when $\zeta_0=0$ ($\omega\neq 0$), if $2\zeta_1-\zeta_2<2\omega$, then the orbits in the phase segment either depart from the viscous matter-dominated, decelerated solution, and end-up at the, also decelerated, scaling solution, or they are repelled from the scaling equilibrium point (corresponding to decelerated expansion always) and are attracted towards the viscous matter-dominated inflationary solution. In either case none of these scenarios is suitable to accommodate the present cosmological paradigm, since there is no any critical point that could be associated with conventional matter and/or radiation dominance, which are included here in the matter component characterized by energy density $\rho_B$.\footnote{As a matter of fact, in the present work, for simplicity, it has been assumed that the conventional matter behaves like pressureless dust, but it is clear that even if consider it to be radiation there would not be any critical point associated with radiation-domination (see the next section).} Radiation and matter-dominated phases are necessary to explain the formation of the amount of cosmic structure we see, in particular the right growth of structure \cite{amendola}.

An interesting situation occurs when $\zeta_2=0$, $\zeta_1\neq0$. In this case, for the viscous matter-dominated solution $x=1$, one has $$q=\frac{1+3(\omega-\zeta_1)}{2}, \qquad \omega_{eff}=\omega-\zeta_1,$$ so that the solution is inflationary if $\zeta_1>(1+3\omega)/3$. Besides, this critical point is a future attractor in the phase space (segment) only if $\zeta_1>\omega$. Hence, this solution is stable and inflationary only if $\zeta_1>\omega+1/3$. Alternatively, the scaling equilibrium point $x=\zeta_1/\omega$, $$\frac{\Omega_B}{\Omega_v}=\frac{\omega-\zeta_1}{\zeta_1}, \qquad q=1/2,\;\omega_{eff}=0,$$ exists whenever, $0\leq\zeta_1\leq\omega$. It is stable whenever it exists, $\zeta_1<\omega$. Hence, when both critical points coexist, the viscous matter-dominated (decelerated) solution is the past attractor, while the (also decelerating) scaling solution is the future attractor. Curiously, if $\zeta_1>\omega+1$, i. e., if the viscous matter-dominated solution is the future attractor (besides, it is the only critical point in the phase segment), the effective EoS parameter behaves like a phantom, $\omega_{eff}<-1$.

If, on the contrary, $\zeta_1=0$, $\zeta_2\neq 0$, then for the viscous matter-dominated critical point $x=1$, $$q=\frac{1+3\omega}{2-3\zeta_2}, \qquad \omega_{eff}=\frac{2\omega+\zeta_2}{2-3\zeta_2},$$ so that this solution corresponds to inflationary expansion if $\zeta_2>2/3$. It is stable when, either, $\zeta_2+2\omega<0$, $\zeta_2<2/3$, or, $\zeta_2+2\omega>0$, $\zeta_2>2/3$. The scaling solution, $x=-\zeta_2/2\omega$, $$\frac{\Omega_B}{\Omega_v}=-\frac{2\omega+\zeta_2}{\zeta_2},$$ exists if $-2\omega\leq\zeta_2<0$. It is stable whenever it exists. In this case the viscous matter-dominated cosmic fluid mimics phantom behavior if $\zeta_2>\omega+1$ ($\omega_{eff}<-1$). As before, no critical point in the phase space can be associated with conventional matter dominance contrary to what is required by the standard cosmological paradigm.

\subsubsection{Case with $\zeta_0\neq0$.}\label{zeta0neq0}

In this case to the already existing variable $x$ one has to add a new one: $$y=\frac{1}{\zeta_0/H+1}\;\Rightarrow\;\frac{\zeta_0}{H}=\frac{1-y}{y}.$$ Hence, the phase space is the bounded plane region $$\Psi=\{(x,y)|0\leq x\leq 1,\;0<y\leq 1\}.$$ As before, the Friedmann constraint reads $\Omega_B=1-x$. The corresponding autonomous system of ODE looks like

\bea &&x'=\frac{3(2-2\zeta_1+\zeta_2)}{(2-3\zeta_2)\;y}(x-1)\left(y+\frac{2\omega x y-2}{2-2\zeta_1+\zeta_2}\right),\nonumber\\
&&y'=\frac{3(2-\zeta_1-\zeta_2)}{2-3\zeta_2}\left(y-1\right)\left(y+\frac{\omega x y-1}{2-\zeta_1-\zeta_2}\right).\label{odexy}\eea For the deceleration and EoS parameters the following expressions are obtained:
\bea q=\frac{(4-3\zeta_1)y+3\omega x y -3}{(2-3\zeta_2)y}, \qquad \omega_{eff}=\frac{(2-2\zeta_1+\zeta_2)y+2\omega x y-2}{(2-3\zeta_2)y}.\label{q_omega}\eea

The critical points of (\ref{odexy}), $P_i:(x_i,y_i)$, together with their main properties are summarized below.

\begin{enumerate}

\item Viscous matter-dominated solution $P_v:(1,1)$. This case corresponds either to the formal limit $\zeta_0=0$, or to the initial singular state characterized by $H\rightarrow\infty$. We have $$q=\frac{1-3\zeta_1+3\omega}{2-3\zeta_2}, \qquad \omega_{eff}=\frac{\zeta_2-2\zeta_1+2\omega}{2-3\zeta_2}.$$ The eigenvalues of the linearization (Jacobian) matrix for this point are:

\bea \lambda_1=\frac{3(1-\zeta_1-\zeta_2+\omega)}{2-3\zeta_2}, \qquad \lambda_2=\frac{3(\zeta_2-2\zeta_1+2\omega)}{2-3\zeta_2}.\nonumber\eea This solution is a past attractor if, either $\zeta_1<\omega+1/3$, $\zeta_2<2/3$, or $\zeta_1>\omega+1/3$, $\zeta_2>2/3$.

\item de Sitter (also viscous fluid-dominated) solution

\bea P_{dS}:\left(1,\frac{1}{2-\zeta_1-\zeta_2+\omega}\right)\;\Rightarrow\;H=H_0=\frac{\zeta_0}{1-\zeta_1-\zeta_2+\omega},\nonumber\eea characterized by $q=-1$, $\omega_{eff}=-1$. The eigenvalues of the Jacobian matrix corresponding to this critical point are:

\bea \lambda_1=-\frac{3(1-\zeta_1-\zeta_2+\omega)}{2-3\zeta_2}, \qquad \lambda_2=-\frac{3}{(2-\zeta_1-\zeta_2+\omega)^2}.\nonumber\eea The de Sitter solution $P_{dS}$ exists whenever $0<y\leq 1$, i. e., if: $\zeta_1+\zeta_2\leq\omega+1$. It is the future attractor in $\Psi$ if: $\zeta_1+\zeta_2<\omega+1$, $\zeta_2<2/3$. Otherwise, if: $\zeta_1+\zeta_2<\omega+1$, $\zeta_2>2/3$, $P_{dS}$ is a saddle critical point instead.

\item Matter/viscous matter-scaling solution, $$P_{B/v}:\left(\frac{2\zeta_1-\zeta_2}{2\omega},1\right), \qquad \frac{\Omega_B}{\Omega_v}=\frac{2\omega-2\zeta_1+\zeta_2}{2\zeta_1-\zeta_2}.$$ This critical point exists ($0\leq x\leq 1$) if, $\zeta_2\leq 2\zeta_1\leq\zeta_2+2\omega$. It is characterized by, $q=1/2$, $\omega_{eff}=0$. The eigenvalues of the corresponding Jacobian matrix are $$\lambda_1=3/2, \qquad \lambda_2=-\frac{3(\zeta_2-2\zeta_1+2\omega)}{2-3\zeta_2},$$ so that, if, $2\zeta_1-\zeta_2<2\omega$, $\zeta_2<2/3$, it is a saddle point in the phase space. Otherwise, if, $2\zeta_1-\zeta_2<2\omega$, $\zeta_2>2/3$, it is the past attractor instead.

\end{enumerate}

The main properties (existence, stability, etc.) are summarized in tables \ref{tab2}, \ref{tab2'}


\begin{table}\centering
\begin{tabular}{| c | c | c | c | c | c | c |}
\multicolumn{6}{c}{\textbf{Critical points ($\omega\neq 0$, $\zeta_0\neq 0$)}}\\
\hline \hline
$P_i$ &  $x$ & $y$& Existence & $\Omega_B$ & $\omega_{eff}$ & q \\
\hline\hline
$P_v$ & $1$ & $1$& Always& $0$& $\frac{2(\omega-\zeta_1)+\zeta_2}{2-3\zeta_2}$& $\frac{1+3(\omega-\zeta_1)}{2-3\zeta_2}$\\
\hline
$P_{dS}$ & $1$ & $\frac{1}{2-\zeta_1-\zeta_2+\omega}$& $\zeta_1+\zeta_2\leq\omega+1$& $0$& $-1$& $-1$\\
\hline
 $P_{B/v}$&  $\frac{2\zeta_1-\zeta_2}{2\omega}$& $1$& $\frac{\zeta_2}{2}\leq\zeta_1\leq\frac{\zeta_2}{2}+\omega$&  $\frac{2(\omega-\zeta_1)+\zeta_2}{2\omega}$& $0$& $1/2$ \\
\hline\hline
\end{tabular}\caption{Existence and relevant parameters of the critical points $P_i$ for the case when $\omega\neq 0$, $\zeta_0\neq 0$.}\label{tab2}
\end{table}

\begin{table}\centering
\begin{tabular}{| c | c | c | c | c | c |}
\multicolumn{6}{c}{\textbf{Stability of the critical points ($\omega\neq 0$, $\zeta_0\neq 0$)}}\\
\hline \hline
$P_i$ &  $x$ & $y$& Stability& $\lambda_1$ & $\lambda_2$ \\
\hline\hline
$P_v$ & $1$ & $1$& Unstable if $\zeta_1<\omega+1/3$, $\zeta_2<2/3$,& $\frac{3(1-\zeta_1-\zeta_2+\omega)}{2-3\zeta_2}$& $\frac{3(\zeta_2-2\zeta_1+2\omega)}{2-3\zeta_2}$\\
 & & & or if $\zeta_1>\omega+1/3$, $\zeta_2>2/3$& & \\
\hline
$P_{dS}$ & $1$ & $\frac{1}{2-\zeta_1-\zeta_2+\omega}$& Stable if $\zeta_1+\zeta_2<\omega+1$, $\zeta_2<2/3$,& $-\frac{3(1-\zeta_1-\zeta_2+\omega)}{2-3\zeta_2}$& $-\frac{3}{(2-\zeta_1-\zeta_2+\omega)^2}$\\
 & & & Saddle if $\zeta_1+\zeta_2<\omega+1$, $\zeta_2>2/3$.& & \\
\hline
 $P_{B/v}$&  $\frac{2\zeta_1-\zeta_2}{2\omega}$& $1$& Saddle if $2\zeta_1-\zeta_2<2\omega$, $\zeta_2<2/3$,& $3/2$& $-\frac{3(\zeta_2-2\zeta_1+2\omega)}{2-3\zeta_2}$\\
 & & & Unstable if $2\zeta_1-\zeta_2<2\omega$, $\zeta_2>2/3$.& & \\
\hline\hline
\end{tabular}\caption{Stability of the critical points $P_i$ for the case when $\omega\neq 0$, $\zeta_0\neq 0$. The eigenvalues of the linearization matrix around a given critical point are $\lambda_1$ and $\lambda_2$.}\label{tab2'}
\end{table}



\begin{figure}[t!]
\begin{center}
\includegraphics[width=7cm,height=7cm]{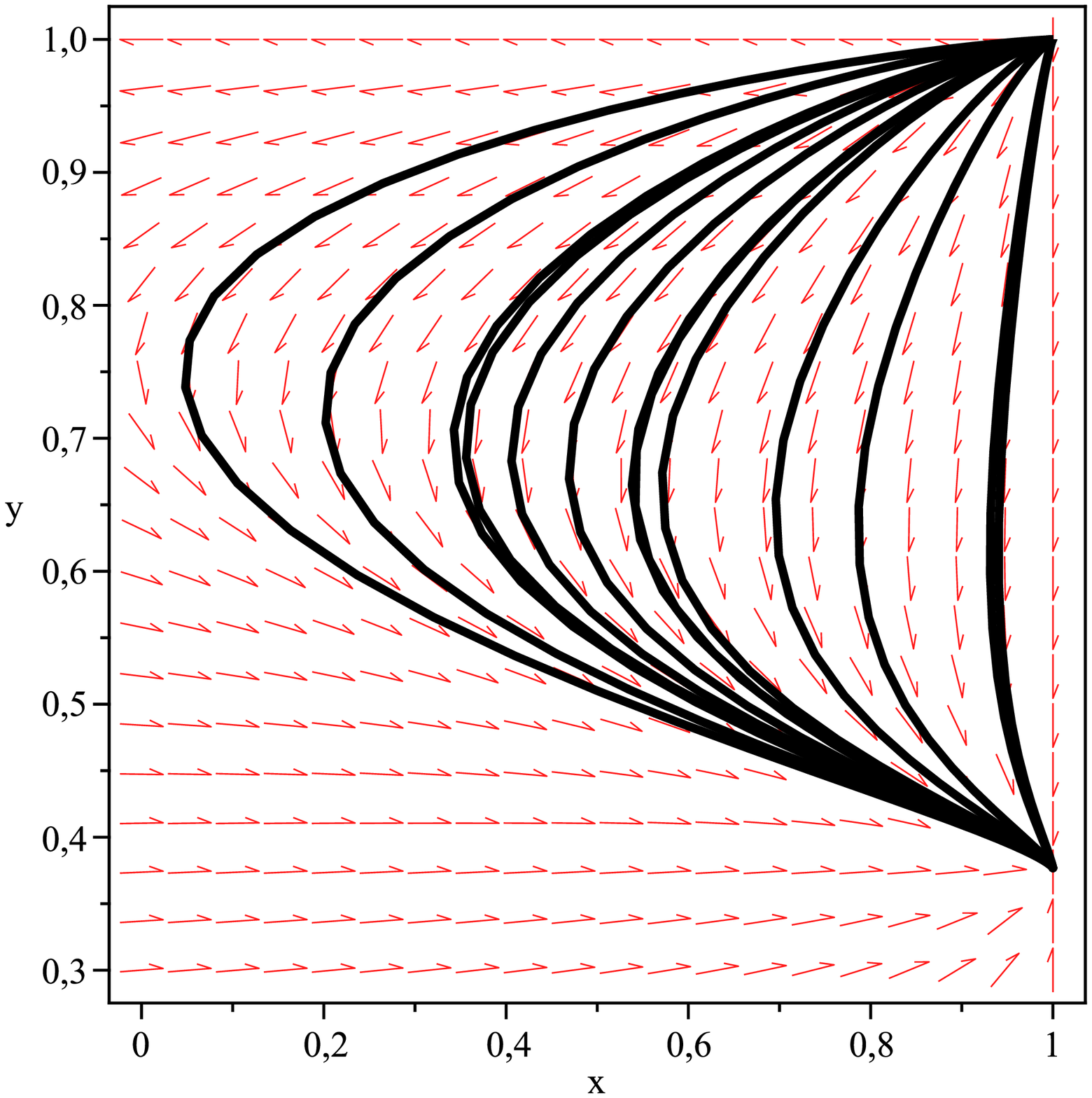}
\includegraphics[width=7cm,height=7cm]{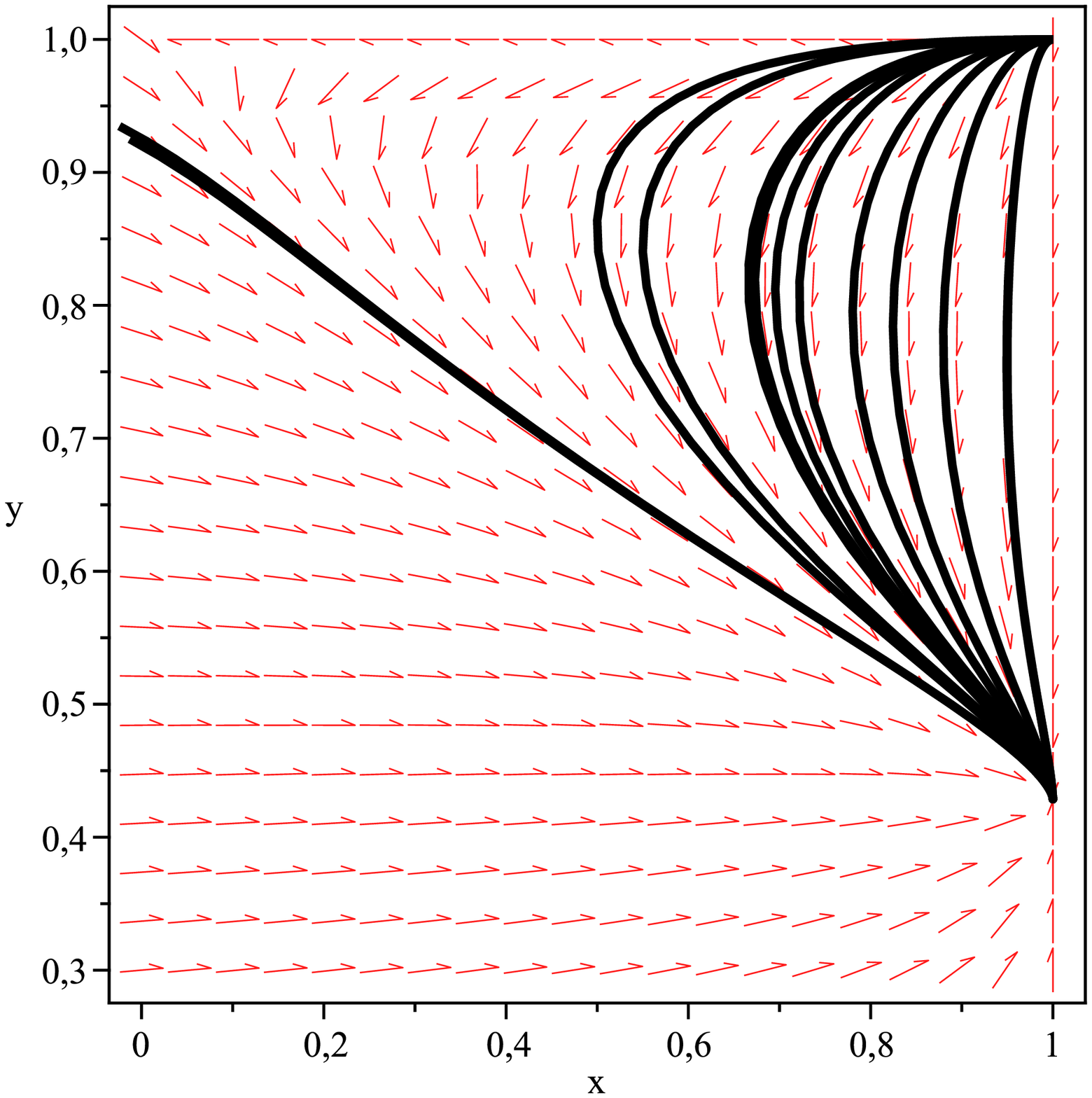}
\caption{Phase portraits $(x,y)$ for the case \ref{zeta0neq0}, for different choices of the free parameters $\zeta_1$, and $\zeta_2$ ($\omega=1/3$). We have taken best estimated values of $\zeta_1=\tilde\zeta_1/3$, and $\zeta_2=6\tilde\zeta_2$ from table \ref{best-estimated}: $\tilde\zeta_1=-0.96$, $\tilde\zeta_2=0$ -- left-hand panel, and $\tilde\zeta_1=0$, $\tilde\zeta_2=0.05$ -- right-hand panel. In both cases the orbits are repelled from the viscous matter-dominated solution and approach to the de Sitter future attractor. The scaling critical point does not exist.} \label{figps1}
\end{center}
\end{figure}


\subsection{Pressureless Viscous Fluid ($\omega=0$).}\label{w=0}

In this case equations (\ref{odexy}) simplify to:

\bea &&x'=\frac{3(2-2\zeta_1+\zeta_2)}{2-3\zeta_2}\left(\frac{x-1}{y}\right)\left(y-\frac{1}{1-\zeta_1+\zeta_2/2}\right),\nonumber\\
&&y'=\frac{3(2-\zeta_1-\zeta_2)}{2-3\zeta_2}\left(y-1\right)\left(y-\frac{1}{2-\zeta_1-\zeta_2}\right).\label{odexyw0}\eea Besides, $$q=\frac{(4-3\zeta_1)y-3}{(2-3\zeta_2)y}, \qquad \omega_{eff}=\frac{(1-\zeta_1+\zeta_2/2)y-1}{2(2-3\zeta_2)y}.$$ Only two critical points of the autonomous system of ODE (\ref{odexyw0}) are found:

\begin{enumerate}

\item Viscous matter-dominated solution $P_v:(1,1)$. The relevant parameters are: $$q=\frac{1-3\zeta_1}{2-3\zeta_2}, \qquad \omega_{eff}=\frac{\zeta_2-2\zeta_1}{2-3\zeta_2},$$ while the eigenvalues of the linearization matrix are: $$\lambda_1=\frac{3(\zeta_2-2\zeta_1)}{2-3\zeta_2}, \qquad \lambda_2=\frac{3(1-\zeta_1-\zeta_2)}{2-3\zeta_2}.$$

\item de Sitter equilibrium point $$P_{dS}:\left(1,\frac{1}{2-\zeta_1-\zeta_2}\right), \qquad q=-1,\;\omega_{eff}=-1,$$ which is also dominated by the bulk viscous matter ($x=1$). The eigenvalues of the Jacobian matrix are: $$\lambda_1=-\frac{3(1-\zeta_1-\zeta_2)}{2-3\zeta_2}, \qquad \lambda_2=-\frac{3}{(2-\zeta_1-\zeta_2)^2}.$$

\end{enumerate}


\begin{table}\centering
\begin{tabular}{| c | c | c | c | c | c | c | c |}
\multicolumn{8}{c}{\textbf{Critical points for the case $\omega=0$ (pressureless viscous fluid)}}\\
\hline \hline
$P_i$ &  $x$ & $y$& Existence & Stability& $\Omega_B$ & $\omega_{eff}$ & q \\
\hline\hline
$P_v$ & $1$ & $1$& Always& Saddle if $\zeta_1+\zeta_2>1$, $2\zeta_1<\zeta_2<2/3$,& $0$& $\frac{\zeta_2-2\zeta_1}{2-3\zeta_2}$& $\frac{1-3\zeta_1}{2-3\zeta_2}$\\
& & & & or if $\zeta_1+\zeta_2<1$, $2/3<\zeta_2<2\zeta_1$,& & & \\
& & & & or if $\zeta_1<1/3$, $\zeta_2<2/3$,&&& \\
& & & & Unstable if $\zeta_1+\zeta_2<1$, $2\zeta_1<\zeta_2<2/3$,& & & \\
& & & & or if $\zeta_1+\zeta_2>1$, $2/3<\zeta_2<2\zeta_1$.&&& \\
\hline
$P_{dS}$& $1$ & $\frac{1}{2-\zeta_1-\zeta_2}$& $\zeta_1+\zeta_2\leq 1$&  Stable if $\zeta_2<2/3$, saddle otherwise& $0$& $-1$& $-1$ \\
\hline\hline
\end{tabular}\caption{Existence, stability and other relevant properties of the critical points $P_i$ for the pressureless viscous fluid case ($\omega=0$).}\label{tab3}
\end{table}

\begin{table}\centering
\begin{tabular}{| c | c | c | c | c | c | c | c |}
\multicolumn{8}{c}{\textbf{Bulk viscous matter with conventional matter and radiation}}\\
\hline \hline
$P_i$ &  $x$ & $y$& Existence & Stability& $\Omega_B$ & $\omega_{eff}$ & q \\
\hline\hline
$P_v$ & $1$ & $1$& Always& Saddle if $\zeta_1+\zeta_2>1$, $2\zeta_1<\zeta_2<2/3$,& $0$& $\frac{\zeta_2-2\zeta_1}{2-3\zeta_2}$& $\frac{1-3\zeta_1}{2-3\zeta_2}$\\
& & & & or if $\zeta_1+\zeta_2<1$, $2/3<\zeta_2<2\zeta_1$,& & & \\
& & & & or if $\zeta_1<1/3$, $\zeta_2<2/3$,&&& \\
& & & & Unstable if $\zeta_1+\zeta_2<1$, $2\zeta_1<\zeta_2<2/3$,& & & \\
& & & & or if $\zeta_1+\zeta_2>1$, $2/3<\zeta_2<2\zeta_1$.&&& \\
\hline
$P_{dS}$& $1$ & $\frac{1}{2-\zeta_1-\zeta_2}$& $\zeta_1+\zeta_2\leq 1$&  Stable if $\zeta_2<2/3$, saddle otherwise& $0$& $-1$& $-1$ \\
\hline\hline
\end{tabular}\caption{Existence, stability and other relevant properties of the critical points $P_i$ for the pressureless viscous fluid case ($\omega=0$).}\label{tab4}
\end{table}


The existence, stability, and other relevant properties of these critical points are shown in the table \ref{tab3}. As seen, as in the former cases, there are not equilibrium points in the phase space that could be correlated with conventional matter-dominance. A phase of conventional matter-dominance is required for the formation of the observed amount of cosmic structure. This is one of the most unwanted features of the viscous fluid scenario and, as shown, this conclusion is irrespective of whether we consider $\omega\neq 0$, or, $\omega=0$. In the next subsection we shall see that this conclusion is robust enough and it holds true even if add a radiation component to the model.

\subsection{Model of bulk viscous matter with conventional matter and radiation}\label{rad}

In this section we shall investigate a more physically involved scenario with bulk viscous matter, with the bulk viscosity coefficient given by (\ref{par_z}). Here, besides a pressureless ($\omega=0$) viscous matter component, we shall include conventional (non-relativistic or dust) matter, and also radiation. The cosmological equations are the following:

\bea &&3H^2=\rho_r+\rho_B+\rho_v, \qquad 6\dot H+6H^2=-2\rho_r-\rho_B-\rho_v+9\zeta H,\nonumber\\
&&\dot\rho_r+4H\rho_r=0, \qquad \dot\rho_B+3H\rho_B=0, \qquad \dot\rho_v+3H\rho_v-9\zeta H^2=0,\label{feqsxyz}\eea where $\rho_r$ is the energy density of the radiation component and, as before, $\rho_B$ and $\rho_v$ stand for the energy densities of non-relativistic (pressureless) matter and of bulk viscous (also pressureless) component, respectively.

In order to transform the above system of equations into a system of autonomous ODE we introduce the following variables of the phase space (the two first variables $x$, and $y$, coincide with the former definitions and we add a new variable $z$):

\be x=\Omega_v, \qquad y=\frac{1}{\zeta_0/H+1}, \qquad z\equiv\Omega_r=\frac{\rho_r}{3H^2}.\label{xyz}\ee The following autonomous system of ODE is obtained,

\bea &&x'=3(1-x)\frac{\zeta}{H}+xz, \qquad y'=\frac{y(y-1)}{2}\left(z+3-3\frac{\zeta}{H}\right), \qquad z'=z\left(z-1-3\frac{\zeta}{H}\right),\label{odexyz}\eea where

\be \frac{\zeta}{H}=\frac{2+\left(2\zeta_1-(z+1)\zeta_2-2\right)y}{(2-3\zeta_2)y}.\label{g}\ee The Friedmann constraint can be written as, $\Omega_B=1-x-z$, while the bounded 3D phase space is given by

\be \Psi=\{(x,y,z)|0\leq x\leq 1,\;0<y\leq 1,\;0\leq z\leq 1\}.\label{psxyz}\ee For the deceleration parameter $q=-1-H'/H$, and the effective EoS parameter $\omega_{eff}=-1-2H'/3H$, one obtains, $$q=\frac{1+z-3\zeta/H}{2}, \qquad \omega_{eff}=\frac{z}{3}-\frac{\zeta}{H},$$ respectively.

Three equilibrium points are found in the phase space $\Psi$ (\ref{psxyz}). A summary of these points, $P_i:(x_i,y_i,z_i)$, together with their main features is given below.

\begin{enumerate}

\item Bulk viscous matter/radiation-scaling
\bea &&P_{r/v}:\left(3(\zeta_2-\zeta_1),\;1,\;1-3(\zeta_2-\zeta_1)\right)\;\Rightarrow\;\frac{\Omega_r}{\Omega_v}=\frac{1-3(\zeta_2-\zeta_1)}{3(\zeta_2-\zeta_1)}.\nonumber\eea This solutions exists if $0\leq 3(\zeta_2-\zeta_1)\leq 1$, i. e., when $\zeta_1\leq\zeta_2\leq\zeta_1+1/3$. In this case the cosmic expansion is decelerating, $q=1$, while $\omega_{eff}=1/3$. The eigenvalues of the Jacobian matrix for this point are, $$\lambda_1=1,\;\lambda_2=2,\;\lambda_3=\frac{2[1-3(\zeta_2-\zeta_1)]}{2-3\zeta_2},$$ so that it is a unstable critical point in $\Psi$ if $\zeta_2<2/3$, and a saddle point otherwise.

\item Bulk viscous matter-dominance, $P_v:(1,1,0)\;\Rightarrow\;\Omega_v=1$. This point is characterized by $$q=\frac{1-3\zeta_1}{2-3\zeta_2}, \qquad \omega_{eff}=\frac{\zeta_2-2\zeta_1}{2-3\zeta_2},$$ and by the following eigenvalues of its linearization matrix:
\bea &&\lambda_1=-\frac{2[1-3(\zeta_2-\zeta_1)]}{2-3\zeta_2},\;\lambda_2=\frac{3(1-\zeta_1-\zeta_2)}{2-3\zeta_2},\;\lambda_3=\frac{3(\zeta_2-2\zeta_1)}{2-3\zeta_2}.\nonumber\eea

\item de Sitter (also bulk viscous matter-dominated, $\Omega_v=1$) solution, $$P_{dS}:\left(1,\;\frac{1}{2-\zeta_1-\zeta_2},\;0\right)\;\Rightarrow\;H=\frac{\zeta_0}{1-\zeta_1-\zeta_2}.$$ For this solution $q=\omega_{eff}=-1$. Since the eigenvalues of the Jacobian matrix are, $$\lambda_1=-4,\;\lambda_2=-3,\;\lambda_3=-\frac{3(1-\zeta_1-\zeta_2)}{2-3\zeta_2},$$ then, if it exists, the de Sitter solution is a stable attractor in $\Psi$ if $\zeta_2<2/3$. Otherwise ($\zeta_2>2/3$) it is a saddle point instead.

\end{enumerate}

We see that a same pattern arises: there are not found conventional dark matter and/or radiation domination critical points in the phase space.\footnote{There is, however, an equilibrium point ($P_{r/v}$) where the radiation and the bulk viscous fluid scale in a constant fraction during the expansion.} As it is suggested by the results of the former subsections, if consider a different EoS parameter $\omega\neq 0$ (say, $\omega=1/3$) for the bulk viscous matter, then an additional conventional matter/bulk viscous fluid-scaling critical point $P_{B/v}$ would arise. In general, there can be found critical points where the different components of the conventional matter (including radiation) scale with the viscous matter, but, in no case conventional matter-dominance is an equilibrium point. This result can rule out the bulk viscous matter-dominated models as acceptable models for the description of the cosmological dynamics of our Universe.


\begin{figure}[t!]
\begin{center}
\includegraphics[width=7cm,height=7cm]{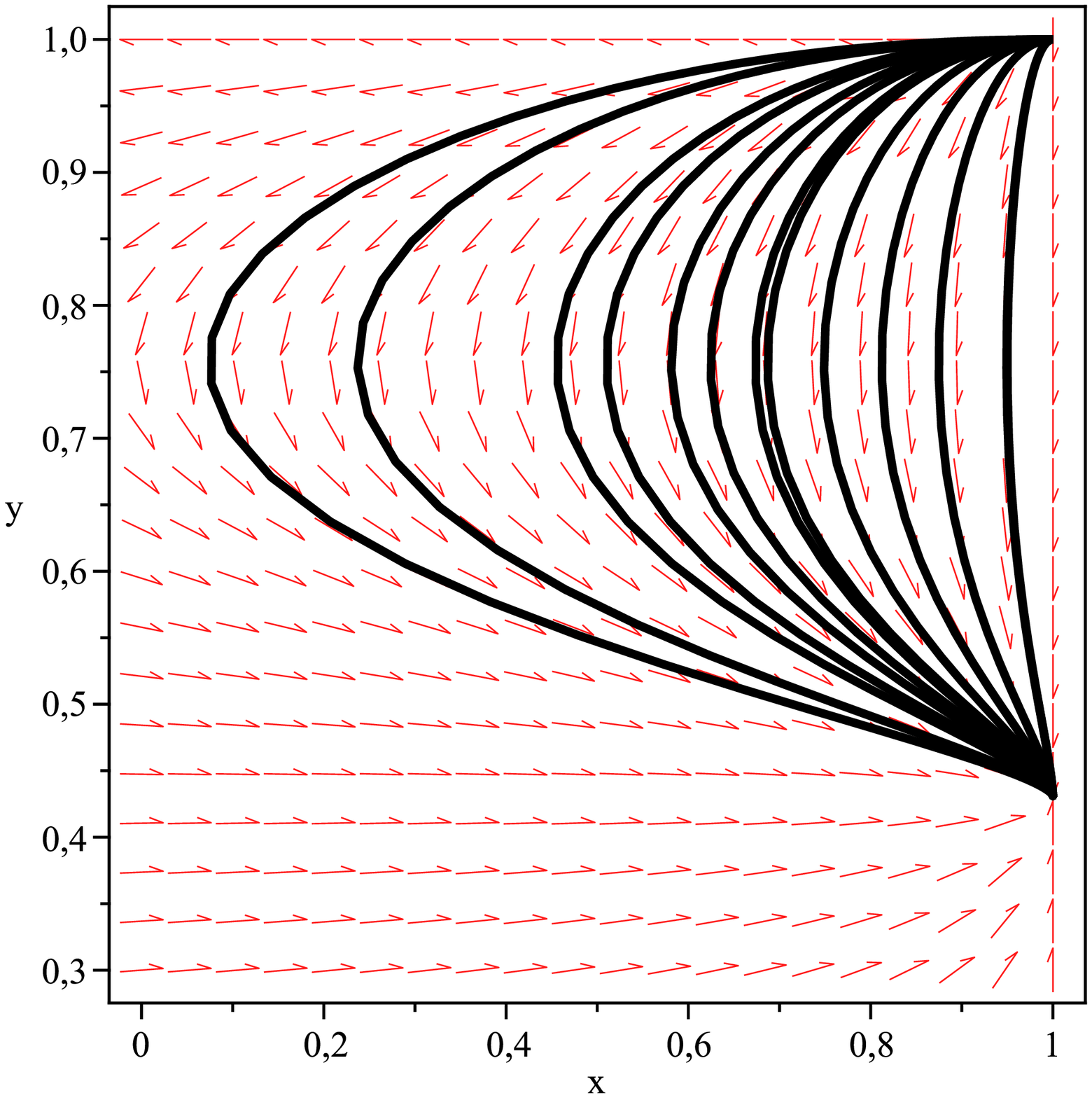}
\includegraphics[width=7cm,height=7cm]{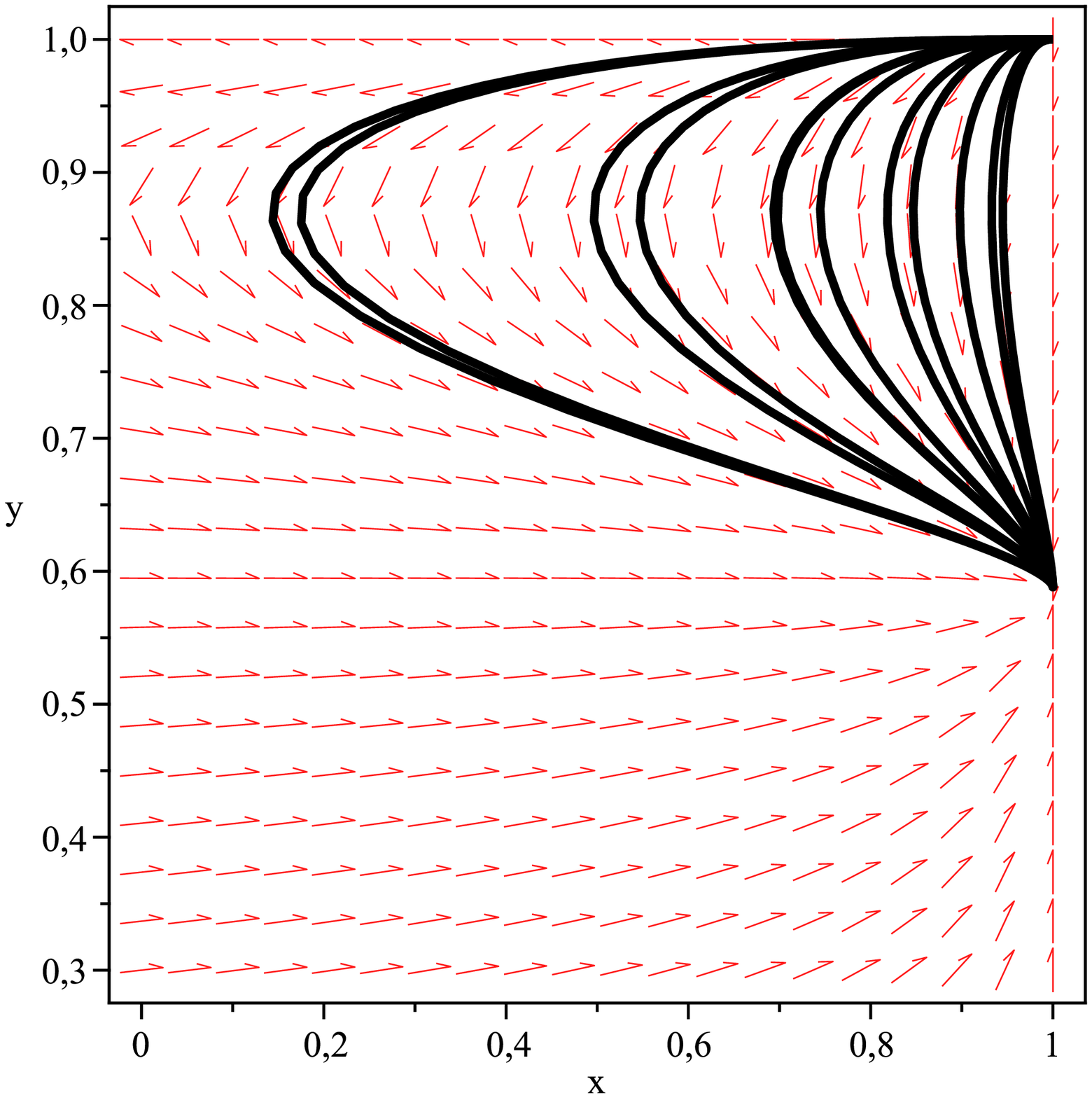}
\caption{Phase portraits $(x,y)$ for the case \ref{w=0}, for different choices of the free parameters $\zeta_1$, and $\zeta_2$ ($\omega=0$).  We have taken the best estimated values of $\zeta_1=\tilde\zeta_1/3$, and $\zeta_2=6\tilde\zeta_2$ ($\tilde\zeta_0\neq 0$) from table \ref{best-estimated}: $\tilde\zeta_1=-0.96$, $\tilde\zeta_2=0$ -- left-hand panel, and $\tilde\zeta_1=0$, $\tilde\zeta_2=0.05$ -- right-hand panel. In both cases the orbits are repelled from the viscous matter-dominated solution and approach to the de Sitter future attractor.} \label{figpsw0}
\end{center}
\end{figure}


\section{Discussion}\label{discussion}

In this section we shall make use of the results of a combined investigation of type Ia supernovae test (section \ref{SNIa}), and of the dynamical systems study (section \ref{d_systems}, see also \ref{app-d-systems}), to check the viability of the bulk viscous matter model to explain the presently accepted cosmological paradigm. Here, without loss of generality, we shall analyze only the model studied in section \ref{rad} where the cosmological dynamics is fueled by radiation, conventional pressureless matter, and bulk viscous (also pressureless) matter. There are found three critical points ($\zeta_0\neq 0$) of the equivalent autonomous system of ODE corresponding to this model (see TAB.\ref{tab4}): i) bulk viscous matter/radiation-scaling solution $P_{v/r}$, ii) bulk viscous matter dominance $P_v$, and iii) de Sitter (also bulk viscous matter dominated) solution $P_{dS}$. If set $\zeta_0=0$ only the bulk viscous matter/radiation-scaling solution (point $P_{v/r}$), and the bulk viscous matter-dominated point $P_v$, survive. In what follows we shall focus in the physically more interesting case where $\zeta_0\neq 0$ which shows a richer phase space dynamics.

There are four regions of interest in the parameters space $(\zeta_1,\zeta_2)$:

\begin{itemize}

\item R1: region where the three points co-exist together. In this region the viscous matter/radiation-scaling solution ($P_{v/r}$) is the past attractor, the bulk viscous matter-dominated phase ($P_v$) is a saddle point, while the de Sitter solution ($P_{dS}$) is the stable attractor.

\item R2: region where only $P_{v/r}$ and $P_v$ co-exist. In this case the bulk viscous matter-dominated solution (point $P_v$) is the past attractor, while, the viscous matter/radiation-scaling solution is a saddle point in the phase space. Not of cosmological interest since there is not critical point which can be associated with a present stage of accelerated expansion.

\item R3: region where only the critical point $P_v$ -- corresponding to viscous matter dominance -- and the de Sitter solution (point $P_{dS}$) co-exist. In this case, either i) $P_v$ is the past attractor and $P_{dS}$ is the stable attractor, or ii) $P_v$ is a saddle point while $P_{dS}$ continues being the future attractor (no past attractor exists), or iii) $P_{dS}$ is a saddle point while $P_v$ is the stable attractor (no past attractor exist). The latter case is not of interest for cosmology since there is not any critical point which could be associated with decelerated expansion.

\item R4: region where the only existing critical point is $P_v$ -- bulk viscous matter dominance. In this region $P_v$ is either a saddle or an unstable critical point. Not of interest for cosmology since there is not any critical point corresponding to the present stage of accelerated expansion.

\end{itemize}

As seen, only regions R1, R3 i), and R3 ii) could be of cosmological interest. If substitute $\zeta_1$ and $\zeta_2$ by their best estimated values in table \ref{best-estimated}, in both cases the region of parameters space R1 above is single out. In this case $P_{v/r}$ is the past attractor, $P_v$ is a saddle point (both $P_{v/r}$ and $P_v$ are associated with decelerated expansion), while $P_{dS}$ is the stable attractor. As mentioned, in R1 there is an equilibrium point (point $P_{v/r}$) where the bulk viscous fluid scales in a constant fraction with radiation. In this case the effective fluid behaves like radiation, but one which is partly bulk viscous. As already mentioned in the introduction, in a cosmological setting the bulk viscosity may arise when the cosmic fluid expands (or contracts) too fast so that the system does not have enough time to restore its local thermodynamic equilibrium and, then, it arises an effective pressure restoring the system to its thermal equilibrium. When the fluid reaches again the thermal equilibrium the bulk viscous pressure vanishes \cite{intro5,intro6}. This would mean that the stage of the expansion associated with $P_{v/r}$ may last for just a brief period of time while the Universe was out of thermodynamic equilibrium, perhaps not enough to produce the right peace of the growth of the fluctuations. If there is a very brief period of radiation domination, then, as long as the fluctuations re-enter the horizon these will not be sufficiently damped, and correspondingly, an unacceptable large value of the dispersion of the density contrast at the scale $8\,h^{-1}$Mpc ($\sigma_8$) might be obtained (see a similar discussion but in the opposite direction in Ref.\cite{amendola}). Besides, in either case although in the bulk viscous matter-dominated stage the effective fluid mimics dust, viscosity may affect the formation of structure in a way that can be observationally tested.

As seen from the above analysis the most serious objection against bulk viscous matter scenarios is the absence of conventional matter and radiation dominated eras. Such a behavior is in marked contradiction with the big bang paradigm according to which, back enough into the past when the temperature of the Universe was larger than $10^4\;K$, the dynamics of the cosmic evolution was driven by a relativistic mixture in the form of radiation (radiation dominated stage). As the Universe expanded and cooled down to temperatures below $10^4\;K$ (and up to $3\times 10^3\;K$), the density of radiation diluted and the cosmic evolution entered in a stage of (non-relativistic) matter dominance. During this phase radiation decoupled from baryons to form the cosmic microwave background and, what is more important, atoms and the derived cosmic structure we see (galaxies, clusters of galaxies, etc.) were formed.

Here we have shown that the absence of such conventional matter and radiation-dominated phases is irrespective of the region in the space of parameters $\zeta_0$, $\zeta_1$, $\zeta_2$, and $\omega$, so that our conclusion is robust enough and the bulk viscous matter-dominated scenario should be ruled out, at least for the parametrization considered in this paper.

\section{Conclusion}\label{conclusion}

In this paper we have applied the dynamical systems tools, in conjunction with the SNe Ia data testing, to judge about the possibility that cosmological bulk viscous matter can stand for an alternative to dark energy. We chose a formerly used parametrization of bulk viscosity \cite{ulises,ulises1} with the addition of a term measuring the influence of the acceleration of expansion: $$\zeta=\zeta_0+\zeta_1 H+\zeta_2\left(\frac{\ddot{a}}{\dot{a}}\right).$$

The study of the asymptotic properties of the model in the equivalent phase space shows that there are not critical points that could be associated with either conventional radiation or matter dominance. This result is independent of the values taken by the free parameters of the model. In consequence, the bulk viscous matter-dominated model is not able to accommodate the presently accepted cosmological paradigm. This argument alone can be considered as a serious objection against cosmological models of bulk viscosity. Notwithstanding, we recommend that other possible parametrizations of bulk viscosity should be considered before concluding to rule out the models.

The authors thank SNI of Mexico for support. A. A. acknowledges the financial support by the Mexican CONACyT and SNI grants 45804 and 56170 respectively and the Instituto Avanzado de Cosmolog\'ia (IAC) collaboration. The work of R. G.-S. was partly supported by SIP20120991, SIP20131811 and also by COFAA-IPN and EDI-IPN grants. U. N. acknowledges financial support from
SNI-CONACYT, PROMEP-SEP and CIC-UMSNH. I. Q. thanks ``Programa PRO-SNI, Universidad de Guadalajara'' for support under grant No 146912.

\section{Acknowledgments}

The authors thank SNI of Mexico for support. A. A. acknowledges the financial support by the Mexican CONACyT and SNI grants 45804 and 56170 respectively and the Instituto Avanzado de Cosmolog\'ia (IAC) collaboration. The work of R. G.-S. was partly supported by SIP20120991, SIP20131811 and also by COFAA-IPN and EDI-IPN grants. U. N. acknowledges financial support from SNI-CONACYT, PROMEP-SEP and CIC-UMSNH. I. Q. thanks ``Programa PRO-SNI, Universidad de Guadalajara" for support under grant No 146912.


\appendix

\section{The viscous density parameter $\tilde{\Omega}_v(z)$}\label{SectionAppendixOmega}
	
In this section we deduce the ordinary differential equation that has to be solved numerically to compute the evolution of the density parameter of the bulk viscosity $\tilde{\Omega}_v(z)$. Inserting equation (\ref{EquivalentFriedmann2Eq}) into (\ref{par_z}), and after re-arranging terms, we obtain the following expression for the  bulk viscosity

\be\zeta=\frac{\zeta_0+\zeta_1 H-\zeta_2\left(\frac{4\pi G}{3H}\right)\Bigl[\rho_B+(1+3w)\rho_v\Bigr]}{1-9\left(\frac{4\pi G}{3}\right)\zeta_2}.\label{ViscosityEq1}\ee

Given the definitions (\ref{DimensionlessViscosityDefinitions}) of the dimensionless bulk viscous coefficients, as well as  $\Omega_{B0}\equiv\rho_{B0}/\rho_{\rm crit}^0$ and
$\tilde{\Omega}_v\equiv\rho_v/\rho_{\rm crit}^0$, the equation (\ref{ViscosityEq1}) can be rewritten in dimensionless form as

\be \tilde{\zeta}=\frac{\tilde{\zeta}_0+\tilde{\zeta}_1 E-\left(9\tilde{\zeta}_2/E\right)\Bigl[\Omega_{B0}a^{-3}+(1+3w)\tilde{\Omega}_v\Bigr]}{9\left(1-9\tilde{\zeta}_2\right)},\label{ViscosityEq1-Dimensionless}\ee with $\tilde{\zeta}_2 \neq 1/9$. Hence, the conservation equation for the viscous component (\ref{cons_eq_vf1}) in terms of dimensionless quantities is the ordinary differential equation  (ODE):

\be a\frac{d \tilde{\Omega}_v}{d a}+ 3(1+w)\tilde{\Omega}_v-9\tilde{\zeta}E=0,\label{ODE1}\ee where $E$ and $\tilde{\zeta}$ are given by the equations (\ref{Friedmann1st-Dimensionless}) and (\ref{ViscosityEq1-Dimensionless}) respectively. Expressing this ODE (\ref{ODE1}) in terms of the redshift $z$ we obtain
\be (1+z) \frac{d \tilde{\Omega}_v(z)}{d z}- 3(1+w)\tilde{\Omega}_v(z)+ 9 \tilde{\zeta}  \left[\tilde{\Omega}_v(z)+\Omega_{B0}(1+z)^3 \right]^{1/2} =0.\label{ODE2}\ee We solve numerically this ODE assuming a dust behavior for the viscous matter (i.e., $w=0$) and the initial conditions $\Omega_{v0} \equiv \tilde{\Omega}_v(z=0) =0.96$, $\Omega_{B0}=0.04$.

The dimensionless Hubble parameter $E \equiv H/H_0$ becomes
\be E(\tilde{\zeta}_0,\tilde{\zeta}_1,\tilde{\zeta}_2)=\sqrt{\tilde{\Omega}_v(z) + \Omega_{B0}(1+z)^3 },\label{ExpEz}\ee where $\tilde{\Omega}_v(z)$ is given by the numerical solution of the ODE (\ref{ODE2}), with  $\tilde{\zeta}_2 \neq 1/9$.

On the other hand, the dimensionless density parameters of the viscous dark matter $\Omega_{v}(z)$ and baryon $\Omega_{m}(z)$ components, can be expressed as
\begin{subequations}
\begin{align} \Omega_{v}(z) &= \frac{\tilde{\Omega}_{v}(z)}{\tilde{\Omega}_{v}(z)+\Omega_{B0}(1+z)^3},\label{Eq-OmegaViscoZ}\\
\Omega_{B}(z) &= \frac{\Omega_{B0}(1+z)^3}{\tilde{\Omega}_{v}(z)+\Omega_{B0}(1+z)^3},\label{Eq-OmegaBaryonZ}
\end{align}\label{Eq-bothOmegasZ}
\end{subequations}  
\noindent where $\tilde{\Omega}_{v}(z)$ is given by the ODE (\ref{ODE2}). The evolution of $\Omega_{v}(z)$ and $\Omega_{B}(z)$ are shown in figure \ref{PlotOmegaZ}. For the deceleration parameter $q(a)=-(\ddot{a}/a)H^{-2}$, one obtains \be q(z) = \frac{\Omega_{B0} (1+z)^3+\tilde{\Omega}_v(z)-9\tilde{\zeta} E(z)}{2(\Omega_{B0} (1+z)^3+\tilde{\Omega}_v(z))},\label{EqqDecelerationZ}\ee where $\tilde{\zeta}$ and $E(z)$ are given by Eqs.(\ref{ViscosityEq1-Dimensionless}) and (\ref{ExpEz}) respectively. The evolution of $q(z)$ is shown in figure \ref{PlotqDecelerationZ}.


\section{Error propagation}\label{SectionErrorPropagation}

In this section we describe the procedure that we followed to compute the propagation of the errors shown in bands of figures \ref{PlotZetaTotal}--\ref{PlotqDecelerationZ}.

Given our ignorance in the possible values for the viscous parameters $(\tilde{\zeta}_0, \tilde{\zeta}_1, \tilde{\zeta}_2)$ before the statistical analysis, we consider  {\it flat} prior probability functions for their values. So, from the Bayes theorem, and assuming that each datum of the SNe and $H(z)$ datasets is Gaussian distributed, the {\it posterior} probability distribution function is proportional to the likelihood function, obtaining

\begin{equation}\label{PosteriorProbabilityFunction}
\text{prob}(\tilde{\zeta}_0,\tilde{\zeta}_1, \tilde{\zeta}_2, | \text{Data}, I) = \text{cte}\cdot\exp\left(- \frac{\chi^2}{2}\right),\end{equation}

where $\chi^2$ is given by the function (\ref{ChiSqrTotal}),
$I$ corresponds to the background information about the parameters and
``cte'' is a normalization constant that contains  our flat prior probability assumptions.

Because of numerical stability it is better to consider the natural logarithm of the posterior probability to work with, i. e., $\mathbf{L} \equiv - \ln [ \text{prob}(\tilde{\zeta}_0, \tilde{\zeta}_1, \tilde{\zeta}_2, | \text{Data}, I) ]$.

Given that the posterior probability is approximately Gaussian near to the best estimated values for the cases $(\tilde{\zeta}_0, \tilde{\zeta}_1)$ and $(\tilde{\zeta}_0, \tilde{\zeta}_2)$ (see figures \ref{plotJustCIZ01} and \ref{plotJustCIZ02}), the covariance matrix $\mathbf{C}$ can be computed as
\begin{equation}\label{CovarianceMatrixDefinition}
 (\mathbf{C}^{-1})_{ij} = \frac{\partial^2 \mathbf{L}}{\partial \tilde{\zeta}_i \partial \tilde{\zeta}_j }
\end{equation}

Next, the variance $\sigma^2$ on any cosmological quantity $Q(\tilde{\zeta}_0, \tilde{\zeta}_1, \tilde{\zeta}_2)$, can be computed with the standard formula for error propagation that takes into account the covariance among the variables (see for instance \cite{BerendsenErrorBook})
\begin{equation}\label{VarianceFormula}
 \sigma^2_Q = \sum^{n}_{i=1} \left(\frac{\partial Q}{\partial \tilde{\zeta}_i} \right)^2 C_{ii} + 2 \sum^{n}_{i=1} \sum^{n}_{j=i+1} \left(\frac{\partial Q}{\partial \tilde{\zeta}_i} \right) \left(\frac{\partial Q}{\partial \tilde{\zeta}_j} \right) C_{ij}
\end{equation}
where $C_{ij}$ corresponds to the $ij$-th element of the covariance matrix $\mathbf{C}$. Given that we take only pairs of parameters then, $n=2$.

Following the prescription indicated in the expression (\ref{CovarianceMatrixDefinition}), we find the following covariance matrices $\mathbf{C}_{(0,1)}$,
$\mathbf{C}_{(0,2)}$ for the cases $(\tilde{\zeta}_0, \tilde{\zeta}_1)$ and $(\tilde{\zeta}_0, \tilde{\zeta}_2)$ respectively,
\begin{equation}\label{CovarianceMatrix}
\mathbf{C}_{(0,1)} =
\begin{pmatrix}
0.1116 & -0.08986  \\
-0.08986 & 0.0730
\end{pmatrix}, \qquad \qquad
\mathbf{C}_{(0,2)} =
\begin{pmatrix}
0.0028 & -0.00039  \\
-0.00039 & 0.000062
\end{pmatrix}.
\end{equation}
For the case $(\tilde{\zeta}_1, \tilde{\zeta}_2)$, it was not possible to find a covariance matrix with real value components.

Using the formula (\ref{VarianceFormula}), the variance expressions for the total bulk viscosity  $\tilde{\zeta}(z)$,
Eq. (\ref{ViscosityEq1-Dimensionless}), the density parameters
$\Omega_{i}(z)$, Eqs. (\ref{Eq-bothOmegasZ}), and the deceleration parameter
$q(z)$, Eq. (\ref{EqqDecelerationZ}), in the case of $(\tilde{\zeta}_0,
\tilde{\zeta}_1)$ as free parameters, have the respective forms
%
\begin{align}
\sigma^2_{\zeta} &= \frac{C_{11}}{324}  \left(
\frac{\tilde{\zeta}_1 \,
   \partial_{(0)}  \tilde{\Omega}_{v}(z)}{\sqrt{(z+1)^3 \Omega_{\rm B0}
+\tilde{\Omega}_v(z)}}+2\right)^2 +  \label{ErrorPropag-TotalViscosity-z0z1} \\
 & \frac{ \left(2 (z+1)^3 \Omega_{\rm B0}+\tilde{\zeta}_1  \, \partial_{(1)}
\tilde{\Omega}_{v}(z) + 2 \tilde{\Omega}_{v}(z) \right)}{324 [(z+1)^3
\Omega_{\rm B0} + \tilde{\Omega}_v(z) ]} \left[ C_{22} \left(2 (z+1)^3
\Omega_{\rm B0}+\tilde{\zeta}_1 \, \partial_{(1)} \tilde{\Omega}_{v}(z) + 2
\tilde{\Omega}_{v}(z) \right) + \right. \nonumber \\
 & + \left.  2 C_{12} \left(2 \sqrt{(z+1)^3 \Omega_{\rm B0} +
\tilde{\Omega}_v(z)} + \tilde{\zeta}_1 \,\partial_{(0)} \tilde{\Omega}_{v}(z) \right)  \right], \nonumber 
\end{align}

\begin{align}
\sigma^2_{\Omega} &= \frac{(z+1)^6 \Omega_{B0}^2  \left(\, \partial_{(0)} \tilde{\Omega}_v(z) (C_{11} \, \partial_{(0)} \tilde{\Omega}_v(z)+2 C_{12} \,
\partial_{(1)} \tilde{\Omega}_v(z))+C_{22} \, \partial_{(1)}
\tilde{\Omega}_v(z)^2\right)}{\left((z+1)^3
\Omega_{B0}+\tilde{\Omega}_v(z)\right)^4}, \label{ErrorPropag-Omega-z0z1} 
\end{align}

\begin{align}
\sigma^2_{q} &= \frac{C_{11} \left(2 (z+1)^3
\Omega_{B0}-\tilde{\zeta}_0 \, \partial_{(0)} \tilde{\Omega}_v(z)+2
\tilde{\Omega}_v(z)\right)^2}{16 \left((z+1)^3 \Omega_{B0}+
\tilde{\Omega}_v(z)\right)^3} + \label{ErrorPropag-q-z0z1} \\
&  + \frac{ C_{12} \left(\frac{\tilde{\zeta}_0
   \, \partial_{(1)} \tilde{\Omega}_v(z)}{\left((z+1)^3 \Omega_{B0}+
   \tilde{\Omega}_v(z)\right)^{3/2}}-2\right)  \left(-2 (z+1)^3
\Omega_{B0}+\tilde{\zeta}_0 \, \partial_{(0)} \tilde{\Omega}_v(z)-2
\tilde{\Omega}_v(z)\right)}{8 \left((z+1)^3 \Omega_{B0}+
   \tilde{\Omega}_v(z)\right)^{3/2}} + \nonumber \\
& + \frac{C_{22}}{16} \left(\frac{\tilde{\zeta}_0 \,
\partial_{(1)} \tilde{\Omega}_v(z)}{\left((z+1)^3
\Omega_{B0}+\tilde{\Omega}_v(z)\right)^{3/2}}-2\right)^2,  \nonumber
\end{align}
\noindent where we use the definition $\partial_{(i)} \equiv \partial / \partial \tilde{\zeta}_i$ and $\tilde{\Omega}_{v}(z)$ is given by the numerical solution of the
differential equation (\ref{ODE2}).
For the case $(\tilde{\zeta}_0, \tilde{\zeta}_2)$ we obtain
\small
%
\begin{align}
\sigma^2_{\zeta} &= \frac{1}{324 (1-9
\tilde{\zeta}_2)^4} \left[4 C_{11} (1-9 \tilde{\zeta}_2)^2
\left(1-\frac{9 \tilde{\zeta}_2
\, \partial_{(0)}  \tilde{\Omega}_v(z)}{2 \sqrt{(z+1)^3
\Omega_{B0}+\tilde{\Omega}_v(z)}}\right)^2 + \right.  \label{ErrorPropag-TotalViscosity-z0z2}   \\
&  \left. 36 C_{12} (1-9 \tilde{\zeta}_2)
\left(1-\frac{9 \tilde{\zeta}_2 \, \partial_{(0)} \tilde{\Omega}_v(z)}{2 \sqrt{(z+1)^3 \Omega_{B0} + \tilde{\Omega}_v(z)}}\right) \left(\frac{-2
(z+1)^3 \Omega_{B0}+\tilde{\zeta}_2 (9 \tilde{\zeta}_2-1) \, \partial_{(2)}
\tilde{\Omega}_v(z)-2 \tilde{\Omega}_v(z)}{\sqrt{(z+1)^3
\Omega_{B0}+\tilde{\Omega}_v(z)}} + 2 \tilde{\zeta}_0\right) + \right.
\nonumber \\
& + \left. 81 C_{22} \left(\frac{-2 (z+1)^3 \Omega_{B0}+\tilde{\zeta}_2 (9
\tilde{\zeta}_2-1) \, \partial_{(2)} \tilde{\Omega}_v(z)-2
\tilde{\Omega}_v(z)}{\sqrt{(z+1)^3 \Omega_{B0}+\tilde{\Omega}_v(z)}}+2
\tilde{\zeta}_0\right)^2 \right], \nonumber
\end{align}

\begin{align}
\sigma^2_{\Omega} &= \frac{(z+1)^6 \Omega_{B0}^2 \left(\, \partial_{(0)}
\tilde{\Omega}_v(z) (C_{11}
   \, \partial_{(0)} \tilde{\Omega}_v(z)+2 C_{12} \, \partial_{(2)}
\tilde{\Omega}_v(z))+C_{22}
   \, \partial_{(2)} \tilde{\Omega}_v(z)^2\right)}{\left((z+1)^3
\Omega_{B0}+\tilde{\Omega}_v(z)\right)^4},  \label{ErrorPropag-Omega-z0z2}
\end{align}

\begin{align}
\sigma^2_{q} &= \frac{C_{11} \left(2 (z+1)^3 \Omega_{B0}-\tilde{\zeta}_0
\, \partial_{(0)} \tilde{\Omega}_v(z) + 2 \tilde{\Omega}_v(z)\right)^2}{16
(1-9 \tilde{\zeta}_2)^2 \left((z+1)^3
\Omega_{B0} + \tilde{\Omega}_v(z)\right)^3} + \label{ErrorPropag-q-z0z2} \\
& + \left\{ 2 C_{12} \left[ 2 (z+1)^3
   \Omega_{B0} - \tilde{\zeta}_0 \, \partial_{(0)} \tilde{\Omega}_v(z)+2
\tilde{\Omega}_v(z)\right] \left[\tilde{\zeta}_0 \left(-18 (z+1)^3
\Omega_{B0}  - 9 \tilde{\zeta}_2 \, \partial_{(2)} \tilde{\Omega}_v(z) - 18
\tilde{\Omega}_v(z) + \right. \right. \right. \nonumber \\
& + \left. \left. \left. \,\partial_{(2)} \tilde{\Omega}_v(z)\right) + 18
\left((z+1)^3 \Omega_{B0} + \tilde{\Omega}_v(z)\right)^{3/2}\right]
\right\} \frac{1}{16(9 \tilde{\zeta}_2-1)^3 \left((z+1)^3
\Omega_{B0} + \tilde{\Omega}_v(z)\right)^3} + \nonumber \\
& +  \frac{C_{22} \left[ 18 \tilde{\zeta}_0 \left((z+1)^3
\Omega_{B0}+\tilde{\Omega}_v(z)\right) + \tilde{\zeta}_0 (9
\tilde{\zeta}_2-1)\, \partial_{(2)} \tilde{\Omega}_v(z) - 18
\left((z+1)^3 \Omega_{B0} + \tilde{\Omega}_v(z)\right)^{3/2}
\right]^2}{16(1 -9 \tilde{\zeta}_2)^4 \left((z+1)^3
\Omega_{B0} + \tilde{\Omega}_v(z)\right)^{3}}. \nonumber
\end{align}

\normalsize

Finally, given the above expressions and the matrix components $C_{ij}$ from (\ref{CovarianceMatrix}), we add and subtract the square root of $\sigma^2_k$ to the central functions (\ref{ViscosityEq1-Dimensionless}), (\ref{Eq-bothOmegasZ}) and  (\ref{EqqDecelerationZ}) respectively for each case to obtain the surrounding boundary lines, the bands, of figures \ref{PlotZetaTotal}--\ref{PlotqDecelerationZ}.

\section{Remarks on phase space analysis}\label{app-d-systems}

Usually the way to test the (theoretical/observational) viability of a given cosmological model is through using known solutions of the cosmological field equations or by seeking for new particular solutions that are physically plausible. However, in general, the cosmological field equations are very difficult to solve exactly and even when an analytic solution can be found it will not be unique but just one in a large set of them. This is not to talk about stability of given solutions.

An alternative way around is to invoke the dynamical systems tools to extract useful information about the asymptotic properties of the model instead. In this regard knowledge of the critical (also equilibrium or fixed) points in the phase space corresponding to a given cosmological model is a very important information since, independent on the initial conditions chosen, the orbits of the corresponding autonomous system of ordinary differential equations (ODE) will always evolve for some time in the neighborhood of these points. Besides, if the point were a global attractor, independent of the initial conditions, the orbits will always be attracted towards it either into the past or into the future. Going back to the original cosmological model, the existence of the critical points can be correlated with generic cosmological solutions that might really decide the fate and/or the origin of the cosmic evolution.

The above interplay between a cosmological model and the corresponding phase space is possible due to an existing isomorphism between exact solutions of the cosmological field equations and points in the equivalent phase space spanned by given variables $(x,y,...)$. When we replace the original field variables $H$, $\rho_{cdm}$, $\rho_{gde}$, etc., by the phase space variables $$x=x(H,\rho_{cdm},...),\;y=y(H,\rho_{cdm},...),\;...,$$ we have to keep in mind that, at the same time, we trade the original set of non-linear second order differential equations in respect to the cosmological time $t$ (cosmological field equations), by a set of first order ordinary differential equations with respect to the variable $\tau=\ln a$: $$x'=f(x,y,...),\;y'=g(x,y,..),$$ etc. The most important feature of the latter autonomous system of ODE is that the functions $f(x,y,...)$, $g(x,y,...)$, $...$, do not depend explicitly on the parameter $\tau$. In other words, we are trading the study of the cosmological dynamics of $H=H(t)$, $\rho_{cdm}=\rho_{cdm}(t)$, $...$, by the study of the flux in $\tau$-parameter of the equivalent autonomous system of ODE. The critical points of this system $P_i:(x_i,y_i,...)$, i. e., the roots of the system of algebraic equations $$f(x,y,...)=0,\;g(x,y,...)=0,\;...,$$ correspond to solutions of the original system of cosmological equations. If consider small linear perturbations around $P_i$ $$x\rightarrow x_i+\delta x(\tau),\;y\rightarrow y_i+\delta x(\tau),\;...,$$ then these would obey the following system of coupled ODE:
\bea \begin{pmatrix} \delta x' \\ \delta y' \\ \vdots \end{pmatrix}=\begin{pmatrix} f_x & f_y & ...\\ g_x & g_y & ...\\ \vdots & \vdots & ...\end{pmatrix}_{P_i}\begin{pmatrix} \delta x \\ \delta y \\ \vdots \end{pmatrix},\label{01}\eea where the square matrix in the right-hand-side (RHS) of (\ref{01}) $J$ is the Jacobian (also linearization) matrix evaluated at $P_i$. If diagonalize $J$ then the coupled system of ODE (\ref{01}) gets decoupled:
\bea \begin{pmatrix} \delta\bar x' \\ \delta\bar y' \\ \vdots \end{pmatrix}=\begin{pmatrix} \lambda_1 & 0 & 0 &...\\ 0 & \lambda_2 & 0 & ...\\ \vdots & \vdots & \vdots & ...\\ 0 & 0 & ... & \lambda_n\end{pmatrix}\begin{pmatrix} \delta\bar x \\ \delta\bar y \\ \vdots \end{pmatrix},\label{02}\eea where $\lambda_1$, $\lambda_2$, etc., are the eigenvalues of the Jacobian matrix $J$, and the linear perturbations $\delta\bar x$, $\delta\bar y$, etc., are linear combinations of $\delta x$, $\delta y$, $...$: $\delta\bar x=c_{11}\delta x+c_{12}\delta y+...$, $\delta\bar y=c_{21}\delta x+c_{22}\delta y+...$, etc. Perturbations in Eq. (\ref{02}) are easily integrated:
\bea \delta\bar x(\tau)=\delta\bar x(0)\,e^{\lambda_1\tau},\;\delta\bar y(\tau)=\delta\bar y(0)\,e^{\lambda_2\tau},\;...\label{03}\eea In case the eigenvalues had non-vanishing imaginary parts the critical point $P_i$ is said to be spiral.\footnote{In general the eigenvalues can be complex numbers.} Depending on the signs of the real parts of the eigenvalues of $J$ the equilibrium point $P_i:(x_i,y_i,...)$ can be classified into:\footnote{In what follows we shall assume the point $P_i$ is an hyperbolic equilibrium point.} i) source point or past attractor if the real parts of all of the eigenvalues were positive quantities, ii) saddle point if at least one of the real parts of the eigenvalues were of a different sign (for example, $Re(\lambda_1)<0$, $Re(\lambda_2)>0$, etc.), and iii) future attractor if the real parts of all of the eigenvalues were negative quantities. In the last case the equilibrium point is stable against small perturbations $\delta x$, $\delta y$, etc., since these exponentially decay in $\tau$-time (see equations (\ref{03})).

If a given equilibrium point $P_a:(x_a,y_a,...)$ were a global attractor, then, independent on the initial conditions chosen $x(\tau_0)=x_0,\;y(\tau_0)=y_0,...$, every orbit in the phase space will approach to $P_a$ into the future ($\{\tau: \tau>\tau_0\}$), i. e., the global (stable) attractor is the end point of any orbit in $\Psi$. On the contrary, if a given critical point $P_s:(x_s,y_s,...)$ were unstable, i. e., small perturbations around $P_s$ uncontrollably grow up with $\tau$, then this point were a past attractor or, also, the source point of any orbit in the phase space. For a third class of critical points, the so called ''saddle points'', depending on the initial conditions chosen, orbits in $\Psi$ can approach to this point, spend some time around it and then be repelled from it to finally approach to the stable attractor if it exists.\footnote{For purposes of space, our discussion here is oversimplified, since, in general, critical points can be of many types, for instance, spiral, etc. Besides, there can be found also (un)stable manifolds such as cycles, etc. To worsen things there can coexist several local attractors, saddle points, etc., so that, in general, a given orbit in the phase space can approach to several saddle points before they end up at a given local attractor.}

Suppose we have a typical phase portrait, composed of a source critical point $P_s$, a saddle point $P_*$, and a stable (global) attractor $P_a$. Each one of these points corresponds to a given solution of the original cosmological equations, $$H=H_s(z),\;H=H_*(z),\;H=H_a(z),$$ respectively. In the above expressions $z$ is the redshift which is related with $\tau$: $\tau=-\ln(z+1)$. A also typical orbit in the phase space will start at $P_s$ for $\tau=-\infty$, then will approach to $P_*$ and, after a finite (perhaps sufficiently long) $\Delta\tau$, will be repelled by $P_*$ to finally be attracted towards $P_a$. The parallel history in terms of the equivalent cosmological dynamics will be the following. The expansion starts with a Hubble parameter dynamics $H=H_s(z)$ then, as the Universe expands, the cosmic history enters a transient period characterized by the dynamics dictated by $H=H_*(z)$. After a perhaps long yet finite period $\Delta z$ the cosmic expansion will abandon the latter phase to enter into a stage which dynamics obeys $H=H_a(z)$ lasting for ever.

To illustrate the above discussion with a concrete cosmological model, take as an example the so called ''lambda-cold-dark-matter'' ($\Lambda$-CDM) model. The cosmological equations for this model in a flat FRW spacetime are,
\bea &&3H^2=\rho_B+\Lambda, \qquad 2\dot H=-\rho_B, \qquad \dot\rho_B+3H\rho_B=0,\nonumber\eea where $\rho_B$ is the energy density of the CDM and $\Lambda$ is the cosmological constant. It is convenient to introduce the variable $x\equiv\Lambda/3H^2$ of the phase segment $\{x|0\leq x\leq 1\}$. The Friedmann constraint can be written as $\Omega_B=1-x$. The autonomous ODE obtained in this case is the following: $x'=3x(1-x)$. There are two critical points of this ODE: i) the matter dominated solution $x=0$ ($\Omega_B=1$), which happens to be a source point (past attractor), and ii) the de Sitter solution $x=1$ ($H=\sqrt{\Lambda/3}$), which is the stable (future) attractor. Hence, since a typical trajectory in the phase segment starts at $x=0$ and end ups at $x=1$, then in this model the cosmic history starts in a matter dominated period (essential for the formation of structure) and end ups in an inflationary stage lasting for ever into the future.

If we improve this model by adding a radiation matter component, there would be three critical points in the (now 2D) phase space: i) a radiation-dominated phase (unstable critical point), followed by ii) a transient matter-dominated stage (saddle point), and iii) a de Sitter point (stable attractor). In this case the cosmic history starts at a radiation-dominated stage, then enters a transient period of matter dominance (essential for the formation of cosmic structure), to finally approach to the stable attractor (the de Sitter phase)
which will last for ever into the future. This is, precisely, the behavior one expects from a model designed to recreate the presently adopted cosmological paradigm. One has to care only about giving appropriate initial conditions $x_0=x(\tau_0)$, etc., so as to get enough formation of structure, i. e., to ensure that the corresponding orbit in the phase space will spend enough time in the neighborhood of the saddle critical point associated with the matter-dominated solution.

We want to underline that, the fact that a given exact solution of the cosmological equations can not be associated with a critical point in the equivalent phase space, does not mean at all that the above solution does not exist. In fact it might exist, but it should not be a stable solution, so it should not be useful in a cosmological setting. To illustrate our point take as an example, again, the $\Lambda$-CDM model. There is a large set of known (classes of) solutions to the cosmological equations of the model, however, only the radiation-dominated, matter-dominated, and the de Sitter solutions, are of importance in a cosmological context which is compatible with the accepted cosmological paradigm. The latter is characterized by the following stages: i) early time inflationary period, followed by ii) a radiation-dominated, and iii) a matter-dominated phases, both associated with a stage of decelerated expansion, and iv) a present period of accelerated expansion which might last forever.


\end{document}